%% file: sigconf.tex
\documentclass[sigconf,nonacm]{acmart}
\AtBeginDocument{%
  \providecommand\BibTeX{{%
    \normalfont B\kern-0.5em{\scshape i\kern-0.25em b}\kern-0.8em\TeX}}}
   
\copyrightyear{2024}
\acmYear{2024}
\setcopyright{acmlicensed}\acmConference[SCID '24]{THE 1st Workshop on Security-Centric Strategies for Combating Information Disorder }{July 2, 2024}{Singapore, Singapore}
\acmBooktitle{THE 1st Workshop on Security-Centric Strategies for Combating Information Disorder (SCID '24), July 2, 2024, Singapore, Singapore}
\acmDOI{10.1145/3660512.3665519}
\acmISBN{979-8-4007-0650-9/24/07}
\usepackage[caption=false]{subfig}
\usepackage{algorithmic}
\usepackage[linesnumbered,ruled,vlined]{algorithm2e}

\newcommand{\framework}{TripletViNet}
\begin{document}

%%
%% The "title" command has an optional parameter,
%% allowing the author to define a "short title" to be used in page headers.
\title{TripletViNet: Mitigating Misinformation Video Spread Across Platforms}

\author{Petar Smolovic}
\authornote{Corresponding authors}
\affiliation{%
  \institution{The University of Sydney}
  \city{Sydney}
  \country{Australia}}
\email{petar.smolovic437@gmail.com}

\author{Thilini Dahanayaka}
\authornotemark[1]
\affiliation{%
  \institution{The University of Sydney}
  \city{Sydney}
  \country{Australia}}
\email{thilini.dahanayaka@sydney.edu.au}

\author{Kanchana Thilakarathna}
\affiliation{%
  \institution{The University of Sydney}
  \city{Sydney}
  \country{Australia}}
\email{kanchana.thilakarathna@sydney.edu.au}

%%
%% The "author" command and its associated commands are used to define
%% the authors and their affiliations.
%% Of note is the shared affiliation of the first two authors, and the
%% "authornote" and "authornotemark" commands
%% used to denote shared contribution to the research.
% \author{Petar Smolovic}
% %\authornote{Both authors contributed equally to this research.}
% %\email{trovato@corporation.com}
% \orcid{1234-5678-9012}
% \author{Thilini Dahanayaka}
% \author{Kanchana Thilakarathna}
% %\authornotemark[1]
% \email{petar.smolovic437@gmail.com}
% \affiliation{%
%   \institution{University of Sydney}
%   \city{Sydney}
%   \state{NSW}
%   \country{Australia}
% }

%%
%% By default, the full list of authors will be used in the page
%% headers. Often, this list is too long, and will overlap
%% other information printed in the page headers. This command allows
%% the author to define a more concise list
%% of authors' names for this purpose.
%\renewcommand{\shortauthors}{Trovato and Tobin, et al.}

%%
%% The abstract is a short summary of the work to be presented in the
%% article.
\begin{abstract}
There has been rampant propagation of fake news and misinformation videos on many platforms lately, and moderation of such content faces many challenges that must be overcome. Recent research has shown the feasibility of identifying video titles from encrypted network traffic within a single platform, for example, within YouTube or Facebook. However, there are no existing methods for cross-platform video recognition, a crucial gap that this works aims to address.  Encrypted video traffic classification within a single platform, that is, classifying the video title of a traffic trace of a video on one platform by training on traffic traces of videos on the same platform, has significant limitations due to the large number of video platforms available to users to upload harmful content to. To attempt to address this limitation, we conduct a feasibility analysis into and attempt to solve the challenge of recognizing videos across multiple platforms by using the traffic traces of videos on one platform only. We propose \framework, a framework that encompasses i) platform-wise pre-processing, ii) an encoder trained utilizing triplet learning for improved accuracy and iii) multiclass classifier for classifying the video title of a traffic trace. To evaluate the performance of \framework, a comprehensive dataset with traffic traces for 100 videos on six major platforms with the potential for spreading misinformation such as YouTube, X, Instagram, Facebook, Rumble, and Tumblr was collected and used to test \framework\ in both closed-set and open-set scenarios. \framework\ achieves significant improvements in accuracy due to the correlation between video traffic and the video's VBR, with impressive final accuracies exceeding 90\% in certain scenarios.

\end{abstract}

%%
%% The code below is generated by the tool at http://dl.acm.org/ccs.cfm.
%% Please copy and paste the code instead of the example below.
%%

% \ccsdesc[500]{Security and privacy~Network security}
% \begin{CCSXML}
% <ccs2012>
%  <concept>
%   <concept_id>00000000.0000000.0000000</concept_id>
%   <concept_desc>Do Not Use This Code, Generate the Correct Terms for Your Paper</concept_desc>
%   <concept_significance>500</concept_significance>
%  </concept>
%  <concept>
%   <concept_id>00000000.00000000.00000000</concept_id>
%   <concept_desc>Do Not Use This Code, Generate the Correct Terms for Your Paper</concept_desc>
%   <concept_significance>300</concept_significance>
%  </concept>
%  <concept>
%   <concept_id>00000000.00000000.00000000</concept_id>
%   <concept_desc>Do Not Use This Code, Generate the Correct Terms for Your Paper</concept_desc>
%   <concept_significance>100</concept_significance>
%  </concept>
%  <concept>
%   <concept_id>00000000.00000000.00000000</concept_id>
%   <concept_desc>Do Not Use This Code, Generate the Correct Terms for Your Paper</concept_desc>
%   <concept_significance>100</concept_significance>
%  </concept>
% </ccs2012>
% \end{CCSXML}

% \ccsdesc[500]{Do Not Use This Code~Generate the Correct Terms for Your Paper}
% \ccsdesc[300]{Do Not Use This Code~Generate the Correct Terms for Your Paper}
% \ccsdesc{Do Not Use This Code~Generate the Correct Terms for Your Paper}
% \ccsdesc[100]{Do Not Use This Code~Generate the Correct Terms for Your Paper}

%%
%% Keywords. The author(s) should pick words that accurately describe
%% the work being presented. Separate the keywords with commas.

%% A "teaser" image appears between the author and affiliation
%% information and the body of the document, and typically spans the
%% page.

%\begin{teaserfigure}
%  \includegraphics[width=\textwidth]{sampleteaser}
%  \caption{Seattle Mariners at Spring Training, 2010.}
%  \Description{Enjoying the baseball game from the third-base
%  seats. Ichiro Suzuki preparing to bat.}
%  \label{fig:teaser}
%\end{teaserfigure}

%\received{20 February 2007}
%\received[revised]{12 March 2009}
%\received[accepted]{5 June 2009}

%%
%% This command processes the author and affiliation and title
%% information and builds the first part of the formatted document.
\maketitle

\input{Sections/Chapter1}

\input{Sections/Chapter2}

\input{Sections/Chapter3}

\input{Sections/Chapter4}

\input{Sections/Chapter5}

\input{Sections/Chapter6}

%% The next two lines define the bibliography style to be used, and
%% the bibliography file.
\bibliographystyle{ACM-Reference-Format}
%%% -*-BibTeX-*-
%%% Do NOT edit. File created by BibTeX with style
%%% ACM-Reference-Format-Journals [18-Jan-2012].

%%
%% If your work has an appendix, this is the place to put it.
\appendix

%-------------------------------------------------------------------------------

\end{document}

%% file: Sections/Chapter1.tex
\section{Introduction} \label{chap:intro}

Video modality has emerged as the preferred medium for disseminating fake news, mis(dis)information, malicious content, etc. primarily for two reasons: (i) human tendency to process audio-visual content more superficially, leading to a higher likelihood of sharing~\cite{sundar2021seeing}, and  (ii) the comparatively greater challenge in detecting doctored or harmful videos than text-based misinformation~\cite{helmus2022artificial}. Content providers have been struggling to moderate the spread of fake audio-visual content as we observed in many recent events such as the COVID pandemic ~\cite{bbcCoronavirusPlandemic}, wars ~\citep{aljazeeraSocialMedia,bbcUkraineWar} and recent elections around the world ~\cite{harvardTwitterFlagged}. 
Despite major content/platform providers having content moderation strategies in place, harmful content can still evade detection. This is primarily because current fake content detection approaches rely on the limitations of the present-day manipulations~\cite{agarwal2020detecting, li2018exposing}. Consequently, these methods  
offer only a short-term solution until the techniques for generating more sophisticated fake content evolve further~\cite{struppek2022learning}.  This results in an arms race with the advantage in favor of undetectable fake content~\cite{agarwal2019limits, helmus2022artificial}. 
%add that Apple NuralHash fails with minimal changes.
As a result, many platforms still rely on manual (human-assisted) moderation which is not scalable in addition to issues of moderator bias. Furthermore, the recent staff layoffs in the tech industry have directly impacted the moderation staff exacerbating the issue of human-assisted moderation~\cite{cnbcTechLayoffs}.

Another key limitation of platform-level moderation is the ability to host the same video on multiple platforms where some may not have content moderation in place, in fact inviting such viral content for financial gain.  For example, the "Plandemic" documentary succinctly highlighted the challenges in stopping the spread of misinformation across platforms ~\cite{bbcCoronavirusPlandemic}. Many misinformation videos that get taken down on mainstream platforms are re-uploaded to 'alternative hosting platforms' with more relaxed moderation guidelines and can continue spreading the content~\cite{migration}.

Therefore, there is an urgent need for a more practical, platform-agnostic method for detecting known fake videos, to supplement platform-level moderation. Detecting the spread of videos at the network level would solve most of the above issues if the challenges of fingerprinting videos from network traffic can be addressed. Such a method can be especially useful in emergency response, mission-critical scenarios and to mitigate threats to national security. 
Unlike the platforms, a network-level moderation process will not have access to the original source video files and hence would need to leverage the statistical properties of passively captured streaming traffic. Moreover, only limited information is visible due to end-to-end encryption. To this end, there has been a significant body of work in recent years in leveraging machine learning models to identify video titles from encrypted network traffic~\citep{schuster2017beauty,li2018deep, Dubin_2017,li2022traffic,khan2021isp,afandi2022fingerprinting,hassan2022youtube,kattadige2021seta++}. Nonetheless, existing work is limited to identifying content in a single platform, i.e. inference is limited to the platform the model is trained on. Given the large number of video platforms available, training models for every platform is not effective or scalable as it would require capturing traffic and training models for each individual platform. Hence it is critical to explore the feasibility of developing a single model that could recognize videos across multiple platforms even when trained on data from a single platform, which has not been studied in the previous work.

In this work, we present the first work that shows the feasibility of cross-platform video identification over encrypted streaming traffic, and unlike the single-platform case, there are no existing methods for baseline comparisons for this task. Rather than naively using one model trained on a given platform on another platform - a strategy we found to be ineffective - we hypothesize that if we can extract the video-specific and platform-specific signatures from a given network traffic trace, we can leverage this information to efficiently identify videos across platforms. This is a challenge that has not been explored in previous work and we propose the framework, \framework, which combines a pre-processor, encoder and classifier in a novel way, to leverage \textbf{triplet learning} to address this. More specifically, we make the following contributions.

\begin{itemize}
    \item We present the first work in cross-platform video recognition over encrypted traffic, trained on a single platform only. We adapt triplet learning to our scenario in a novel way, building triplets using multiple platforms (or "datasets") where one platform ("dataset") provides the anchor and another platform(s) provides the positives and negatives for the triplets.
    \item We conduct an extensive set of experiments to collect a valuable dataset of encrypted traffic of streaming 100 videos on six different platforms such as YouTube, Facebook, X, Instagram, Tumblr, and Rumble, some of which have never been explored in the network traffic fingerprinting literature, and make it publicly available ~\cite{sourcecode}. We also develop a baseline traffic pattern for every video with the standard Variable Bitrate (VBR) patterns.
    \item We utilize the collected data to evaluate the proposed framework that achieves about three times higher accuracy on average using triplet learning in a closed-set and can achieve high accuracies up to 98\% for some platform pairs. We also show the robustness of our framework in the more challenging open-set setting, achieving up to about 80\% known class precision for some platform pairs. 
    \item We investigate and compare various input data variations, classifiers, and parameter variations in our cross-platform video recognition scenario such as binary classification and augmentation, and also show that our method has some robustness to modifications to the videos.
\end{itemize}

The remainder of this paper is structured as follows. Section~\ref{chap:related} presents the related work and Section~\ref{subsec:scenario} and Section~\ref{chap:chap4} introduce our framework for cross-platform video recognition and the experimental setting respectively. Section~\ref{chap:results} presents the results and Section~\ref{chap:conclusion} concludes our work.

%% file: Sections/Chapter2.tex
%-------------------------------------------------------------------------------
\section{Background and Related works}\label{chap:related}

%-------------------------------------------------------------------------------

Encryption was meant to defend the confidentiality of data being transmitted, preventing attackers from eavesdropping and finding out what users were doing~\cite{afandi2022fingerprinting}, such as which video they are streaming. However, recent works have shown that the metadata and statistical properties of encrypted network traffic can still expose certain properties of the data of users~\cite{li2022traffic}, including which video they are streaming~\cite{Dubin_2017}. This is possible because of the observable content-specific temporal patterns due to characteristics of video streaming protocols such as the widely popular “DASH” (Dynamic Adaptive Streaming over HTTPS) protocol~\cite{yang2021clustering}.
The key properties that contribute to such a temporal pattern are; (i) the variable bitrate (VBR) encoding of the video, and (ii) the streamed video data being split into chunks or segments of fixed video playback duration. Scenes in the video with higher action than other scenes require more bits to encode, resulting in a video-specific VBR pattern, i.e. the segments have different sizes corresponding to the content of the video thus, leaking information about the video being streamed despite the encryption~\cite{schuster2017beauty}. It is also worth mentioning that different video platforms have different segmentation parameters which can change over time ~\cite{schuster2017beauty}. However, this pattern of variable size bursts can still be used to achieve high accuracy in video title classification from encrypted video streams~\cite{gu2018walls}. 

\subsection{Encrypted video traffic classification}

Encrypted video traffic classification has gained significant interest recently leading to many works showing the feasibility of leveraging a variety of openly accessible attributes such as burst sizes~\cite{Dubin_2017}, packet directions~\cite{li2018deep}, packet sizes and packets per time interval~\cite{kattadige2021seta++, madarasingha2022videotrain++}, segment sizes of the video streams~\citep{bjorklund2023see}. For example, one of the early works~\citet{schuster2017beauty} tried to identify the video being streamed from four different platforms which are Vimeo, YouTube, Amazon Prime, and Netflix, and they achieved high accuracies above 90\% on all four platforms by creating feature vectors using the bytes per peak, bytes per second and other features such as packet lengths. Attempts to classify the YouTube video of an encrypted video stream by matching the streams to the Variable Bitrate (VBR) pattern of source video files, achieving 74\% using a neural network binary classifier, is considered to have relatively high performance but is inadequate for open-set classification~\cite{schuster2017beauty}. 

Another work, ~\citet{reed2017identifying}, automatically obtained video segment sizes from the Netflix website for 200 videos, and achieved an impressive accuracy of 99.5\% using just a few minutes of the streams ~\cite{reed2017identifying}.
More recently, ~\citet{bjorklund2023see} directly obtained the segment sizes of videos from the server for the SVT Play video platform to build video fingerprints and used a similarity matching algorithm to achieve 98\% accuracy. In summary, these methods have demonstrated the feasibility of network-level moderation of fake video content for a given platform.

\emph{However, none of the existing literature explored the relationship between the encrypted video streams of the same video between different platforms. Thus, in this paper, we investigate the existence of content-specific VBR patterns across platforms and then propose the first work on a cross-platform video traffic identification framework.}

\subsection{Triplet learning for feature embedding}

Triplet loss is one of the most popular loss functions used in metric learning and is used in triplet learning to generate an embedding such that data samples of the same class are closer together in the embedding space and samples of different classes are further away in the embedding space, according to a distance function such as the Euclidean distance~\cite{towardsdatascienceTripletLossAdvanced}. Triplet loss was first introduced mainly in the field of image classification, especially facial recognition, as a way to achieve higher performance ~\cite{wang2014, facenet}. One of the very first works to introduce triplet loss was~\citet{wang2014}, which utilized online triplet mining to obtain fine-grained image similarity and rank the similarity of images ~\cite{wang2014}. Triplet learning was also famously used in FaceNet~\cite{facenet} for face verification and recognition, where facial images were transformed into 128-byte vectors, which could then be used for KNN classification. 

The first work to propose triplet learning for traffic fingerprinting used it for website traffic fingerprinting, where it demonstrated improvement over traditional approaches of transfer learning for few-shot learning with a KNN~\cite{web2019}. They compared different parameter variations and methods for classification and also used triplet learning with data of mutually exclusive sets of class labels between training and testing sets, using 775 classes (websites) for training and triplet learning, and using 100 other classes, even with different distributions, for testing, and demonstrated that the triplet learning method is still fairly effective under such scenarios~\cite{web2019}. Some other works used triplet learning in similar scenarios.~\citet{firstchannel} used triplet learning in side-channel attacks in cryptographic systems, utilizing online semi-hard triplet mining to achieve competitive accuracy for a traditional template attack compared to purely deep-learning-based side-channel attacks. Another work~\cite{power}, tried to use triplet learning to classify the encryption key of power traces, achieving competitive accuracies, needing far fewer samples than the benchmark CNN and with higher robustness ~\cite{power}. 

\emph{Inspired by the performance of triplet learning in the above applications, in this paper, we adopt triplet learning to extract the video-specific signatures embedded in traffic patterns of two different streaming services.}

%% file: Sections/Chapter3.tex
\section{Cross-platform Video Identification}
%-------------------------------------------------------------------------------
\subsection{Threat model and Challenges}\label{subsec:scenario}

\begin{figure}[h!]
    \centering
    \includegraphics[width=1\linewidth]{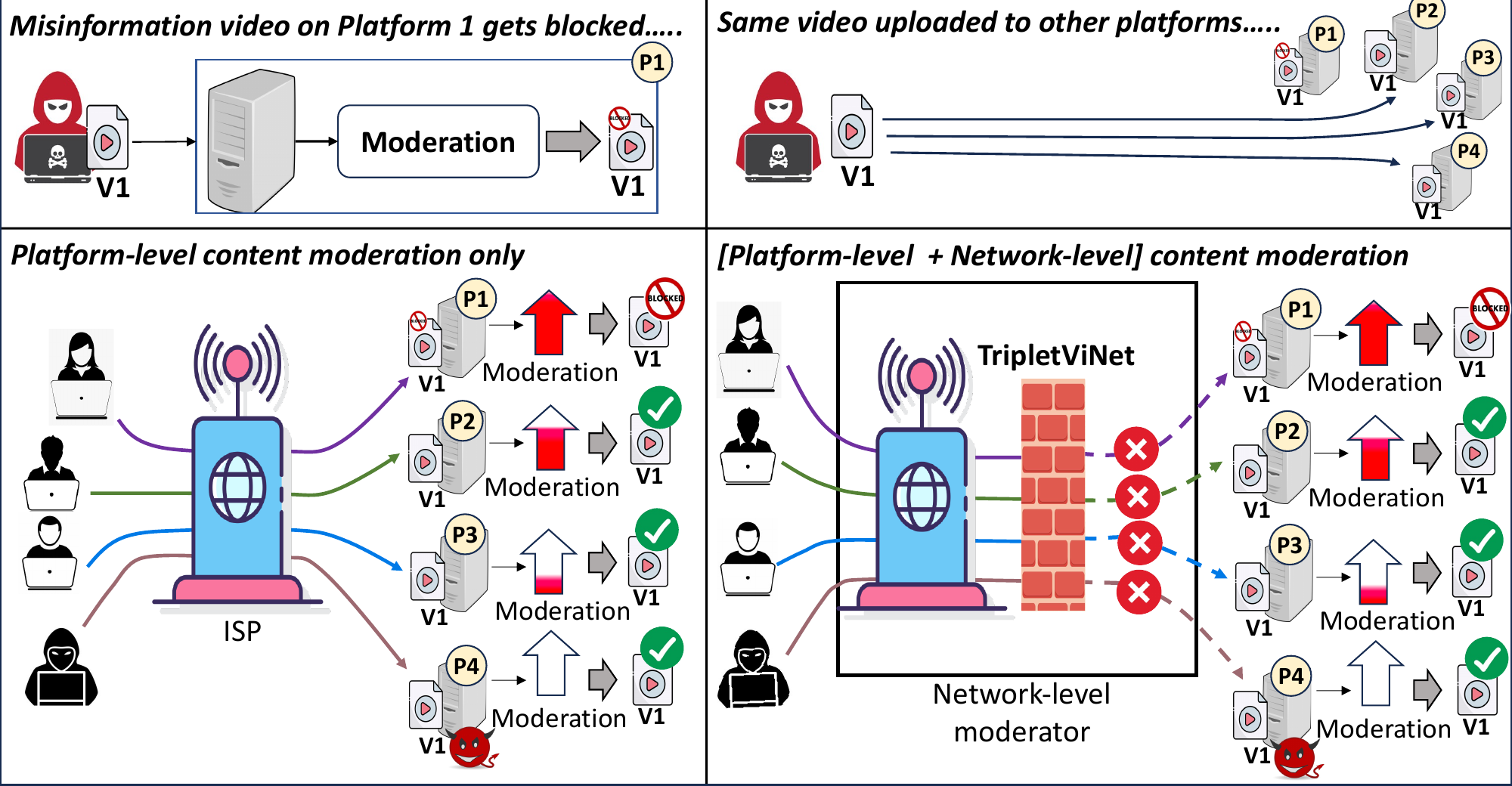}
    \caption{Scenario: Content moderation}
    \label{fig:threatmodel}
\end{figure}

\textbf{Scenario:} We illustrate an example scenario where an adversary uses multiple platforms to evade content filtering in Figure~\ref{fig:threatmodel}. Here the adversary first uploads \textit{V1}, a misinformation video to Platform 1 (P1), but it gets blocked by the moderation process at P1. The adversary retaliates by uploading the same video to multiple other platforms such as P2, P3, P4, etc. as shown in the top right of Figure~\ref{fig:threatmodel}. These different platforms would have varying levels of content moderation or none at all and hence they cannot be guaranteed to identify \textit{V1} as malicious. Additionally, some of the platforms might be malicious themselves and support the spread of \textit{V1}. As shown in the bottom left of Figure~\ref{fig:threatmodel}, until the moderation processes at these platforms flag \textit{V1} as harmful, the content will be freely propagated through social media, despite being blocked by P1. As a solution, we propose network-level content moderation to augment the platform-level content moderation as shown in the bottom right half of Figure~\ref{fig:threatmodel}. We hypothesize that a single machine learning model can be generated to detect the video title irrespective of the served platform observing data traffic flow. Trained machine learning models for malicious videos can be shared across platforms and ISPs (Internet Service Providers) similar to maintaining a blacklist of hashes in platform moderation.
To this end, the network providers can contribute to blocking the spread of harmful content flows at the network. ISPs can offer content filtering as a middle-box service to customers. More importantly, this blocks the spreading of malicious content via non-cooperative content providers. 
%But, an ISP that is aware that \textit{V1} was blocked by Facebook can use machine learning models to compare encrypted traffic signatures of streaming \textit{V1} over Facebook against encrypted traffic flows of its users and identify which users are streaming \textit{V1} (e.g. \textit{A} is watching \textit{V1} over Instagram) and block these connections.

%\subsection{Challenges in recognising videos across platforms}\label{subsec:challenges}

\begin{figure}[!h]
    \centering
    \includegraphics[width=1\linewidth]{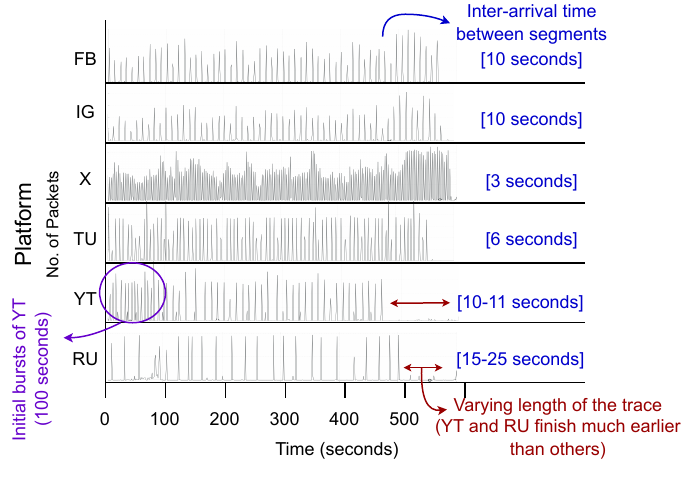}\vspace{-2mm}
    \caption{Comparison of video traffic for the same video on the 6 video platforms}
    \label{figure__5}
\end{figure}

\begin{figure*}[!h] 
\centering
  \subfloat[Facebook (FB)]{% 
    \includegraphics[width=0.14\textwidth]{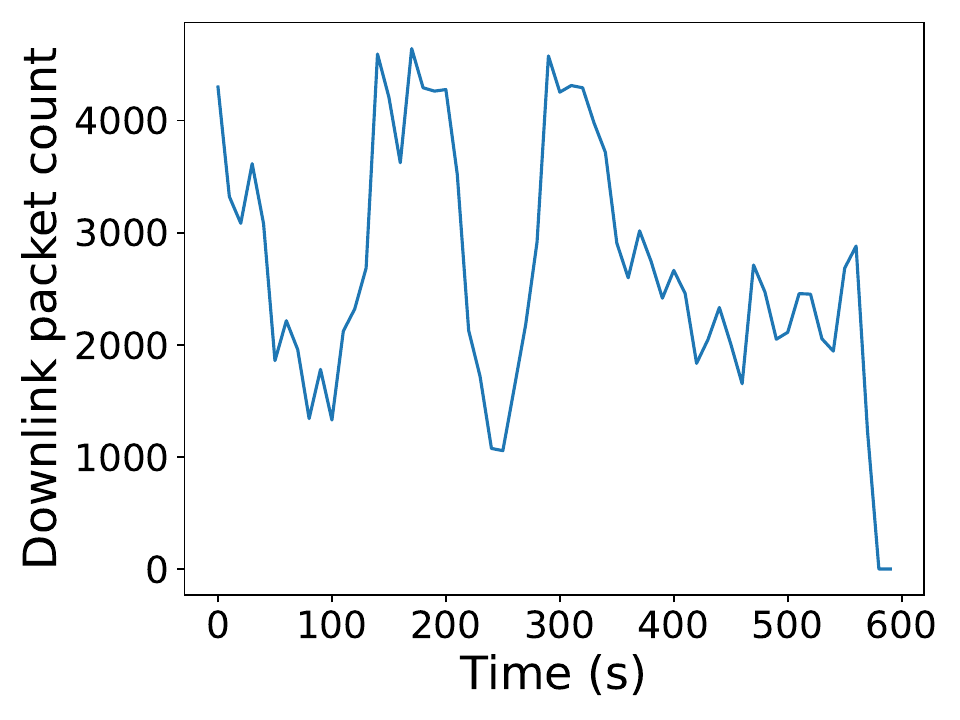}
    \label{fig:Figure4_a}
  } 
    \subfloat[Instagram (IG)]{% 
    \includegraphics[width=0.14\textwidth]{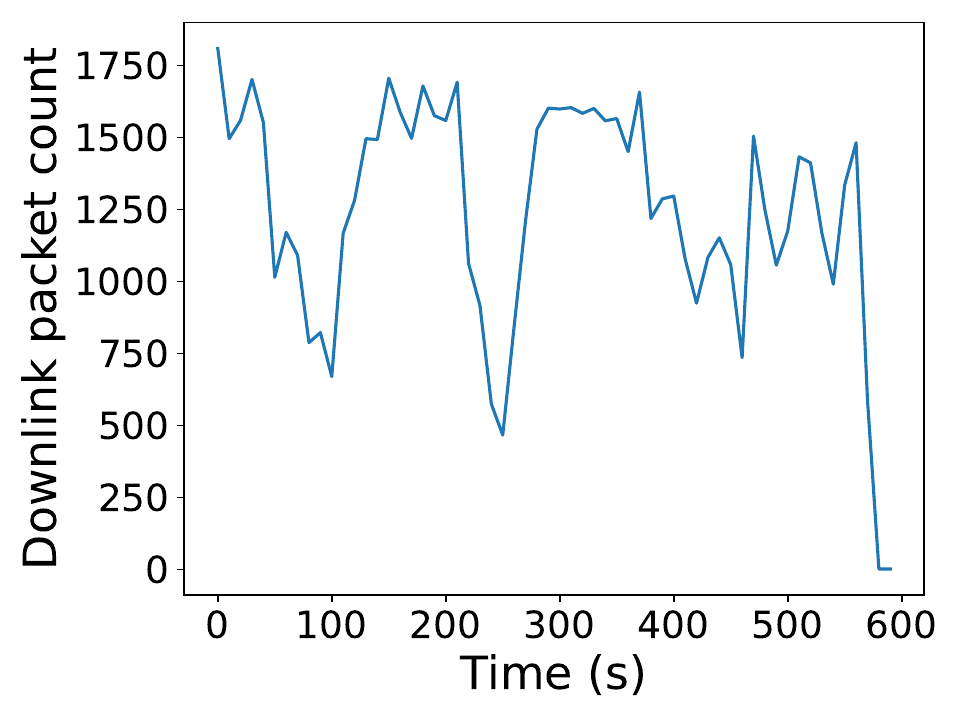}
    \label{fig:Figure4_b}
  } 
  \subfloat[Rumble (RU)]{% 
    \includegraphics[width=0.14\textwidth]{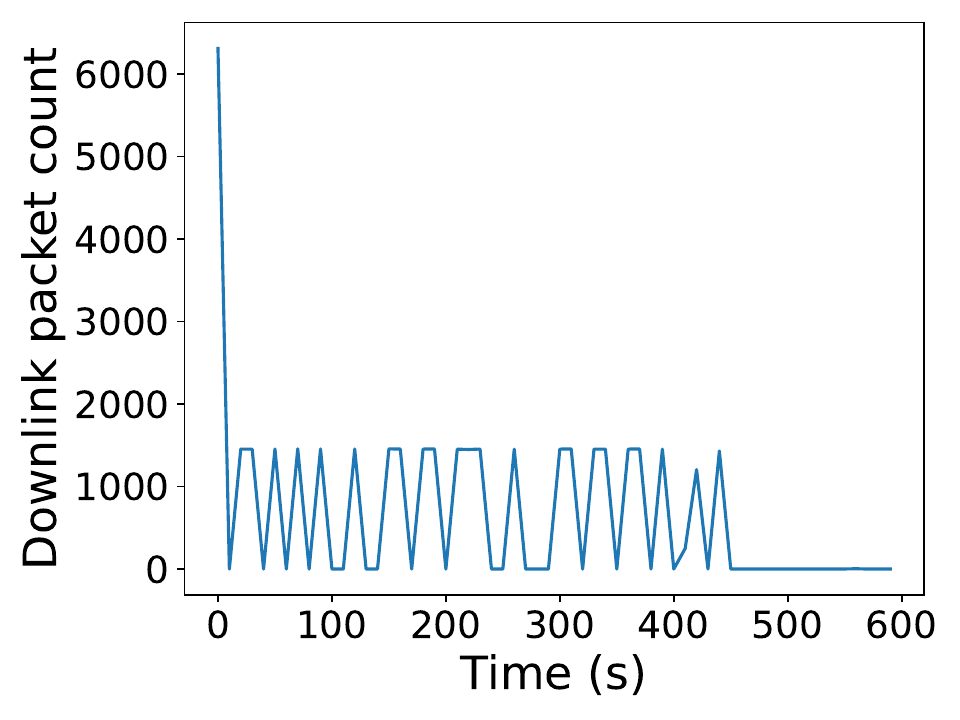}
    \label{fig:Figure4_c}
  }
  \subfloat[Youtube (YT)]{% 
    \includegraphics[width=0.14\textwidth]{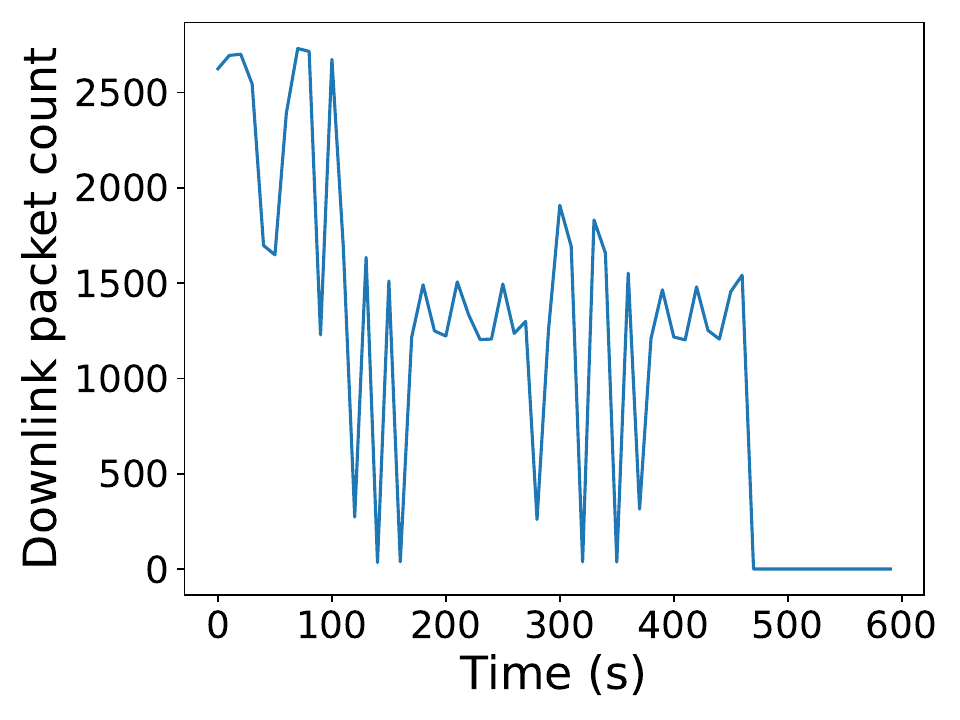}
    \label{fig:Figure4_d}
  }
  \subfloat[Tumblr (TU)]{% 
    \includegraphics[width=0.14\textwidth]{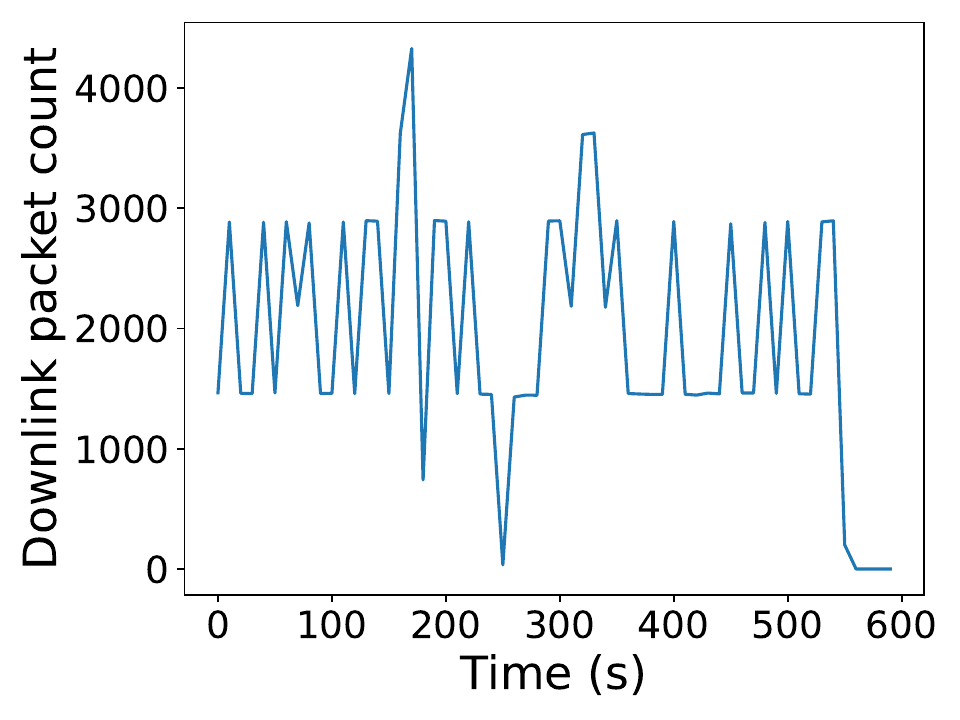}
    \label{fig:Figure4_e}
  }
  \subfloat[X]{% 
    \includegraphics[width=0.14\textwidth]{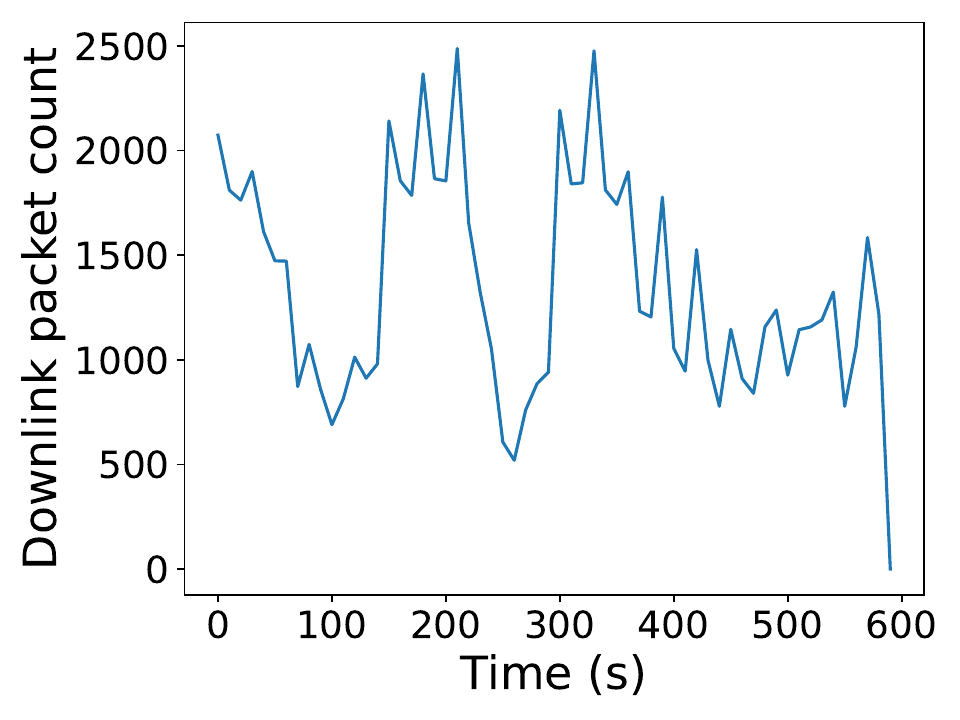}
    \label{fig:Figure4_f}
  }
  \subfloat[VBR]{% 
    \includegraphics[width=0.14\textwidth]{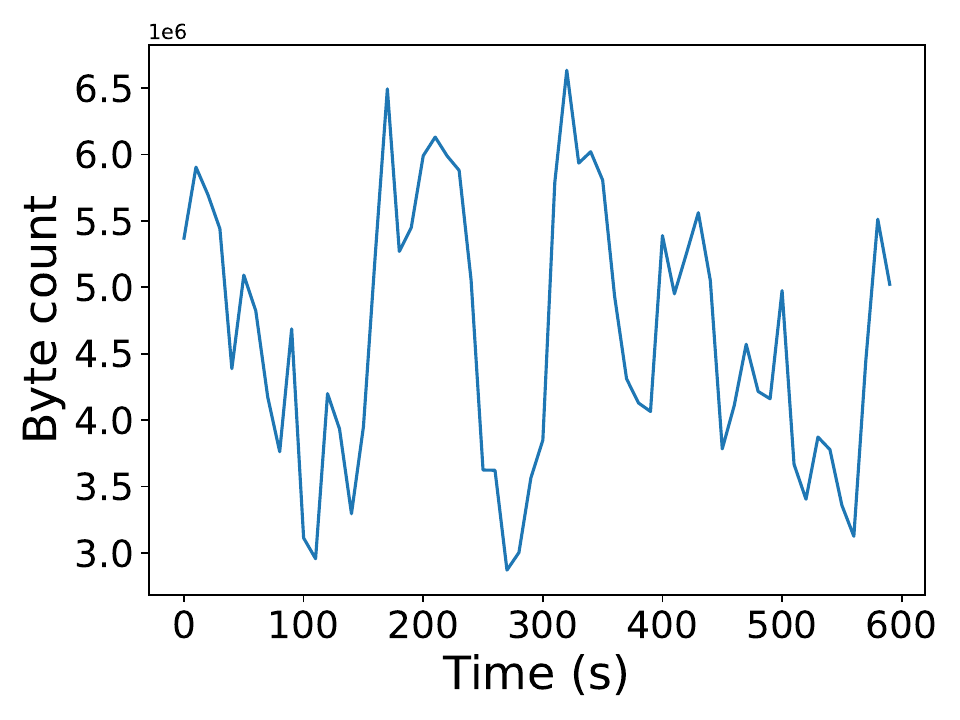}
    \label{fig:Figure4_g}
  }
  \caption{Traffic from 1 run of the same video (with ID 1) using a bin size of 10 seconds} \vspace{-2mm}
  \label{fig:Figure4}
\end{figure*}

\noindent\textbf{Challenges:} The fundamental challenge in the identification of misinformation videos is the end-to-end encryption of streaming traffic. If there is no encryption, conventional DPI (Deep Packet Inspection) tools can be easily utilized to identify the video title. Thus, we will have to rely on the statistical properties of traffic patterns. Figure ~\ref{figure__5} illustrates traffic patterns of the same video for all six platforms, Youtube (YT), Facebook (FB), Instagram (IG), X, Tumblr (TU), and Rumble (RU). We can see that the segmentation parameters such as the time interval between when the client receives each video chunk varies from 3 seconds to 25 seconds between platforms. %For FB and IG it is about 10 seconds, for X it is about 3 seconds, for Tumblr it is about 6 seconds, for Youtube it is about 10-11 seconds and for Rumble it varies between about 15-25 seconds. 
In addition, YT is unique in that it has an initial buffering period of $~$100 seconds where segments are received at about two times the presentation time. The size of segments also varies across platforms leading to varying lengths of transmission even for the same video. For example, YT and RU effectively end sooner than for the other platforms. However, it is also important to mention that the segmentation parameters of the platforms can change over time ~\cite{schuster2017beauty}. There is also a variation in network protocol usage between platforms for sending the downlink video traffic packets with YT using the QUIC protocol, FB and IG using the UDP protocol, and the other 3 platforms (RU, TU, X) using the TCP protocol. Thus, it is not straightforward to extract content-specific signatures out of traffic patterns due to the domination of platform-specific characteristics.

Figure ~\ref{fig:Figure4} shows the traffic pattern of the same video for a relatively large bin size of 10 seconds. While many platforms still have noticeably different patterns, we see that FB, IG, and X share great similarities (even Tumblr is reasonably similar) due to high VBR-dependence, which suggests that the same video has a similar traffic pattern across platforms. We also obtained the variable bitrate (VBR) encoding pattern for the same video using \texttt{ffprobe} of FFmpeg~\cite{ffmpegFfprobeDocumentation} for comparison. The VBR pattern shows high similarity to traffic from FB, IG, and X showing the existence of video-specific patterns resulting from high VBR-dependence, which we hypothesise can be used to recognise videos across platforms.

However, YT and RU are quite different, because the platform-specific features are more dominant than the hidden video-specific features, that is, their traffic has lower VBR-dependence. %so we might need some method to separate out video-specific features from platform-specific features. 

In the remainder of this section, we present how we address these challenges by first incorporating tailored pre-processing techniques in \S~\ref{subsec:preprocess}, followed by pair-vise relationship learning inspired encoding method in \S~\ref{subsec:triplet}, and finally multiclass classifier for detecting video titles in \S~\ref{subsec:multiclass}.

\subsection{\framework: Cross-Platform Video Recognizer using Triplet Learning}

\begin{figure*}[h!]
    \centering
    \includegraphics[width=1\linewidth]{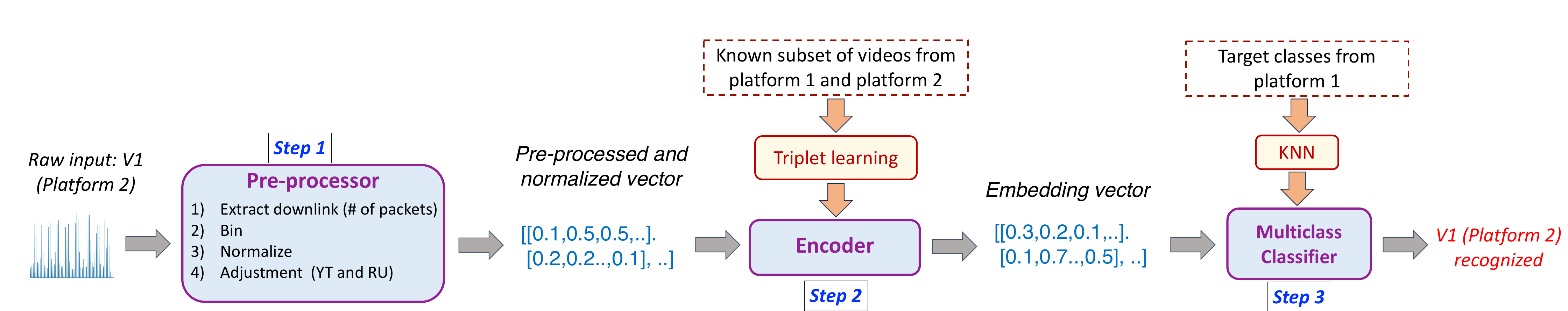}
    \caption{Overview of \framework}
    \label{fig:pipeline}
\end{figure*}

We propose Cross-platform Triplet Video Recognizer (\framework), a framework for recognizing videos across platforms, and address the challenges discussed in Section~\ref{subsec:scenario}. Figure~\ref{fig:pipeline} shows an overview of the framework with three main components \textit{Pre-processor}, \textit{Encoder}, and the \textit{Multiclass Classifier} and each component is explained in detail in the following sections.

\subsubsection{Pre-processor}\label{subsec:preprocess}
\framework\ uses the 'number of packets on the downlink' feature from the captured traffic as this feature leads to the highest performance as per our experiments similar to~\cite{li2018deep}. Accordingly, the pre-processor first extracts the number of packets in the downlink and then bins the trace to 10 second bins (see Section~\ref{chap:hyperpara}), before applying min-max normalization. 

In Section~\ref{subsec:scenario}, we observed that traffic captures from YT and RU have considerable traffic bursts in the beginning and shorter trace lengths compared to other platforms. With this intention of minimizing the effect from such platform-specific characteristics on recognizing videos across platforms, we further pre-process traffic from YT and RU. Accordingly, we extend the initial burst of the capture with the aim of removing the bursty nature at the beginning of the trace and having similar trace lengths as other platforms. More specifically, the first 100 seconds of a traffic capture from YT is extended to 200 seconds with half the amplitude, and the first 520 seconds of a RU traffic capture is extended to 600 seconds using skimage's resize function ~\cite{scikitimageRescaleResize}. We heuristically experimented with pre-processing techniques and found the above methods to be most effective in boosting the overall performance of the framework.

\subsubsection{Encoder}\label{subsec:triplet}
In Section~\ref{subsec:scenario}, we observed the traffic patterns for streaming the same video on some platforms show similarities while traffic patterns for platforms such as YT and RU show significant variations. While the pre-processing steps discussed in Section~\ref{subsec:preprocess} can reduce the effect of platform-specific features to a certain extent, it is not sufficient to achieve satisfactory performance in cross-platform video recognition, which implies the necessity for an improved methodology to more effectively extract video-specific features from streaming traffic. Hence, we explore the feasibility of leveraging \textit{Triplet Learning} ~\cite{towardsdatascienceTripletLossAdvanced} towards this requirement.

\begin{figure}[!h]
    \centering
    \includegraphics[width=1
\linewidth]{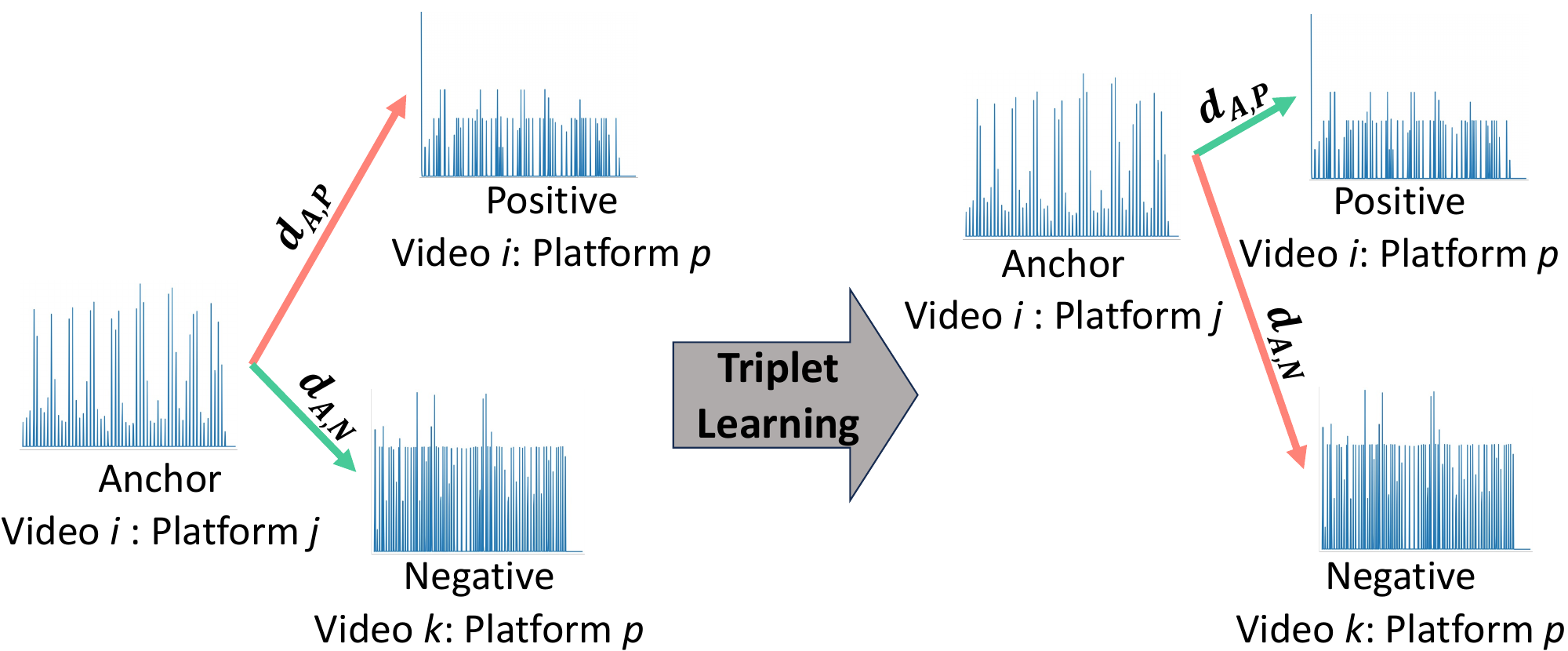}
    \caption{Triplet learning}
    \label{fig:triplet-learning}
\end{figure}

Triplet learning leverages \textit{triplet loss} to train a machine learning model that can encode raw input samples into an embedding space where samples of the same class are clustered together while samples from different classes are pushed further apart. We hypothesize that triplet learning can be adapted to create an embedding space where traffic traces of the same video are clustered together irrespective of the platform it was streamed on, while traces of different videos are pushed further apart as shown in Figure~\ref{fig:triplet-learning}. Accordingly, the resulting embedding space can be considered to represent a traffic trace based on its video-specific characteristics only. Towards this goal, we define triplet loss as given in Equation~\ref{eq:triplet_loss}, where the \textit{anchor} ($A_{ij}$) is a traffic trace of a given video $V_i$ from platform $P_j$, the \textit{positive} ($P_{i\tilde{j}}$) is a trace from $V_i$ on a different platform ($j\neq\tilde{j}$) and the negative ($N_{k\tilde{j}}$) is a trace of a different video ($i\neq{k}$) from the same platform as ($P_{i\tilde{j}}$). Here, $d$ represents the \textit{euclidean distance} and $\alpha$ represents a pre-defined margin.

\begin{equation} \label{eq:triplet_loss}
Loss = max(d(A_{ij}, P_{i\tilde{j}}) - d(A_{ij}, N_{k\tilde{j}}) + \alpha, 0) 
\end{equation} 

\begin{figure}[h!] 
\centering
  \subfloat[Raw input]{% 
    \includegraphics[width=0.23\textwidth]{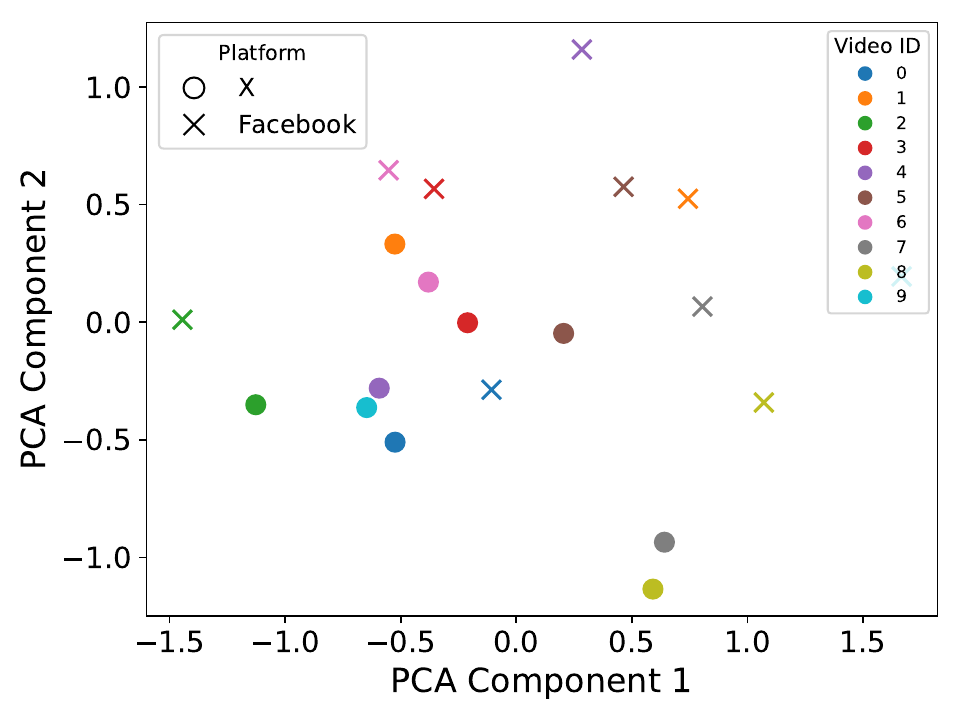}
    \label{fig:raw_input_pca}
  } 
    \subfloat[After triplet learning]{% 
    \includegraphics[width=0.23\textwidth]{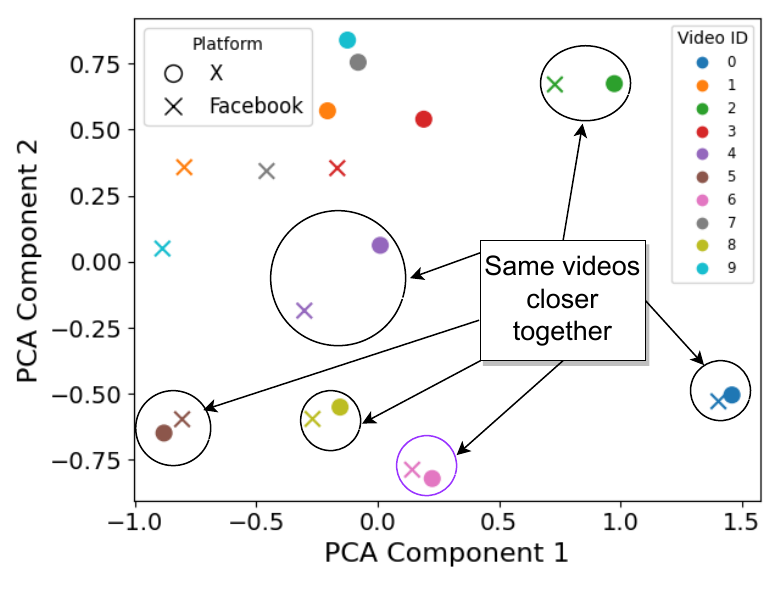}
    \label{fig:triplet_pca}
  } 

  \caption{Distance between classes with PCA analysis} \vspace{-2mm}
  \label{fig:grad_cam}
\end{figure}

To validate our hypothesis, we first randomly choose 10 video titles, and for each title, we select a sample from FB and X and plot the raw inputs in Figure~\ref{fig:raw_input_pca} using PCA analysis. Next, we train an encoder using triplet loss on samples from FB and X and, in Figure~\ref{fig:triplet_pca} we plot the embeddings from the encoder for the same set of samples used in Figure~\ref{fig:raw_input_pca} using PCA. When comparing Figures~\ref{fig:raw_input_pca} and ~\ref{fig:triplet_pca}, we observe that raw samples of the same video from the two platforms are generally randomly placed and are far apart (Figure~\ref{fig:raw_input_pca}), while the embeddings from the encoder for samples of the same video on both platforms are generally placed close together in Figure~\ref{fig:triplet_pca}. Accordingly, this observation validates our hypothesis that triplet learning can be leveraged to extract video-specific signatures from streaming traffic and hence can improve the performance of cross-platform video recognition.

Based on the above observations, we propose a triplet-learning-based encoder that extracts video-specific features from streaming traffic as the second component of \framework. More specifically, the encoder is a deep neural network consisting of a base model trained with triplet loss (Equation~\ref{eq:triplet_loss}) on a known, labeled subset of videos from two platforms. We tested various neural network architectures (Figure \ref{fig:Figure10_a}) and chose the best performing LSTM as the base network in our framework. Furthermore, we leverage non-random offline triplet mining where we generate triplets ensuring every anchor is paired with one instance of every possible negative class.

We tested and compared three base model neural network architectures for the triplet model (encoder) which included: an MLP with 3 dense layers (128 neurons each) and 0.1 dropout layer; a 1D CNN with a convolutional layer (8 filters, size 3) followed by 1D max-pooling (size 2), a Flatten layer, 2 dense layers (128 neurons each), and a 0.1 Dropout layer; and an RNN using just the CuDNNLSTM architecture of Tensorflow and a 0.1 Dropout layer.

\subsubsection{Multiclass Classifier}\label{subsec:multiclass}
As the third component of \framework, the goal of the multiclass classifier is to recognize video titles using the embedding vectors (with video-specific features) provided by the encoder. 
The multi-class classifier is trained on data from a given single platform to classify target (misinformation) videos on that platform. However, as the encoder produces an embedding where the video-specific features of a sample are dominant, the same classifier can identify target videos on a different target platform without being trained on the target platform. In \framework, we use a KNN model as the main multi-class classifier. 

We also explored using a CNN classifier for multiclass classification (see Section~\ref{chap:classifiers} for comparison with KNN) and additionally tested a CNN binary classifier (see section~\ref{chap:binary}) as additional experiments. The binary classifier we built takes as input one sample from 1 platform and one sample from a second platform (or VBR) and outputs whether they are the same video. We were inspired by ~\citet{schuster2017beauty} who use a binary classifier that takes in an input of a video file VBR and a traffic sample and tries to classify whether the VBR pattern and the traffic sample come from the same video, achieving 74\% accuracy which is reasonably good but not suitable for open-set classification. We deal with class imbalance by randomly undersampling the majority class (a pair of samples come from different videos) to balance classes in the training and testing sets that have mutually exclusive videos/classes. We train the binary classifier with the same set of videos used in the triplet model classifier with the mutually exclusive remaining set of videos and compare performance before and after triplet learning (embedding generation) for the training and testing data. 

For both the binary and multiclass CNN classifiers, we used a 1D CNN architecture with: a convolutional layer (8 filters, size 3), a 1D max-pooling layer (size 2), a Flatten layer, 2 dense layers (128 neurons each), a Dropout layer (rate 0.1), and a final Softmax layer. The Softmax layer has 2 neurons for the binary classifier, 20 neurons for the closed set, and 11 neurons for the open set. The models were trained with a learning rate of 0.1, using the SGD optimizer and the "sparse\_categorical\_crossentropy" loss function ~\cite{sparse}.

%% file: Sections/Chapter4.tex
\section{Experiments} \label{chap:chap4}
In this section, we first discuss the data collection process and then discuss the implementation details of our evaluation setup.

\subsection{Data Collection}

We recorded and uploaded 100 of our own 10-minute videos to the considered six platforms, as private or unlisted videos. Then, we collected the video streaming traffic data in a wired network using an HP Envy laptop across 6 months (between April and October of 2023) using \textit{tshark} ~\cite{tsharkk} to record the traffic as PCAP files, \textit{Selenium Webdriver} ~\cite{selenium} and web crawling code to automatically play the videos. We made sure that we cleared the browser cache before starting a new video. Our final dataset consisted of 20 trials or traces per video class, each 10 minutes in duration, for the same 100 videos each uploaded to all 6 platforms. This resulted in roughly 12,000 traces and about ~3 months' worth of non-stop video watching which equals to about a terabyte of data, which is a substantial amount of data that can provide statistically significant results. Table~\ref{dataset} provides the summary of the dataset.

\begin{table}[!h]
\begin{tabular}{|p{2.6cm}|p{5.3
cm}|} \hline
Feature & Description \\ \hline
Number of videos           & 100                     \\ \hline
Trials per video          & 20      \\ \hline
Platforms         & 6 (FB, IG, X, RU, TU, YT)          \\ \hline
Duration per video            & \verb|~| 10 minutes          \\ \hline
Traffic capture used            & 10 minutes                  \\ \hline
Video types                 & 68 live-action, 32 animated    \\ \hline
Data collection          & Wired network, HP Envy laptop, Selenium Webdriver ~\cite{selenium}, tshark ~\cite{tsharkk}      \\ \hline
Collection period          & 6 months (April to October 2023)      \\ \hline
Total data used         & 12,000 traces, \verb|~| 3 months of video, \verb|~| 1TB      \\ \hline
Video quality          &  Diverse (representative of the real world)     \\ \hline
Video privacy status          & Private/unlisted (no ads encountered)      \\ \hline
\end{tabular}
\caption{Summary of dataset}  
\vspace{-3mm}
\label{dataset}
\end{table}

Note that the use of public videos was considered but due to several reasons including the rarity of finding the same video on many platforms, the uncertainty as to whether seemingly similar videos on different platforms are the same video, and more advanced encoding for popular videos ~\citep{schuster2017beauty,fbb}, we chose to upload and test our own videos instead. Furthermore, using private and unlisted videos also allowed us to conduct these experiments without extra permission and external ethics approvals. 68 videos (about 2/3) were live-action videos in a variety of real-life environments, and 32 (about 1/3) were animated videos from screen recordings of online gameplays, giving us a comprehensive variety of videos and making them more representative of the real world. We do not use any specific video quality, which is also more representative of the real world. Moreover, as we conducted our experiments over 6 months we captured data under different network conditions reflecting a realistic dataset. 

We also obtained the VBR from the video files using \textit{ffprobe} of \textit{FFmpeg}~\cite{ffmpegFfprobeDocumentation}, using various segment durations (corresponding to bin sizes), and got the sizes of segment files from video segmentation files in a video segmentation process similar to what many video streaming platforms do.

\subsection{Evaluation setup}

We used Tensorflow and Keras to implement the neural networks and triplet learning model and used other libraries and packages like Sklearn for additional machine learning and other implementation components. The tests were conducted in a Google Colab environment with access to GPUs and reading data from Google Drive folders. 

We evaluate the performance of \framework\ in closed-set and open-set scenarios. The main closed-set experiment was with 80 videos/classes to train the encoder and a mutually exclusive set of 20 videos/classes to test an additional classifier. Additionally, we conducted many other experiments such as with parameter variations and varying the number of classes and traces per class to train the encoder, and also using 10 classes for classification (with 90 classes to train the encoder) rather than 20 as an additional test. We conducted the open-set test with 80 videos/classes to train the encoder, 10 known videos/classes to train a CNN classifier, and then 10 additional classes/videos in an 'unknown class' combined with the 10 known classes (50/50 split) to test the open set classifier. All the tests were run with cross-validation to ensure robustness and stability of results and eliminate bias. We rotate the testing set in each scenario so that for each scenario, each fold of cross-validation has a mutually exclusive testing set, and all 100 videos are included in a testing set. For example, we use 5-fold cross-validation for 20 videos closed set scenario and 10-fold cross-validation for 10 videos closed set scenario. We also randomly select the classes/videos in the testing and training sets with a consistent random seed to avoid bias. 

We conducted extensive tests with refined parameters and settings from empirical experimentation and used various base models (CNN, MLP, RNN) for the encoder as well as various techniques and classifiers with specific configurations and architectures. For the closed-set scenario, we only focus on mean accuracy rather than other metrics such as precision and recall, because the classes are balanced and accuracy gives an intuitive and general overview of model performance. In contrast, for the open-set scenario, we focus on mean precision for trained videos to minimize false positives which can be important to avoid censorship issues. The processing speed for training the pre-processor was reasonably fast, taking between a few seconds to a few minutes depending on the input bin size. Encoding and classification were both very fast, each taking only up to a few seconds.

We use 5 epochs and 128 batch size for all training of the encoder when using non-random offline triplet mining, and use 20 epochs and 128 batch size when using online semi-hard triplet mining. The embeddings generated were all 128 neurons as this worked best. All the encoders used the 'Adam' optimizer and triplet loss.

%-------------------------------------------------------------------------------

%% file: Sections/Chapter5.tex
%-------------------------------------------------------------------------------
\section{Results}\label{chap:results}
%-------------------------------------------------------------------------------
\subsection{\framework\ classification performance: Closed set}

\begin{figure*}[!h] 
\centering
  \subfloat[Platform pairs]{% 
    \includegraphics[width=0.19\textwidth]{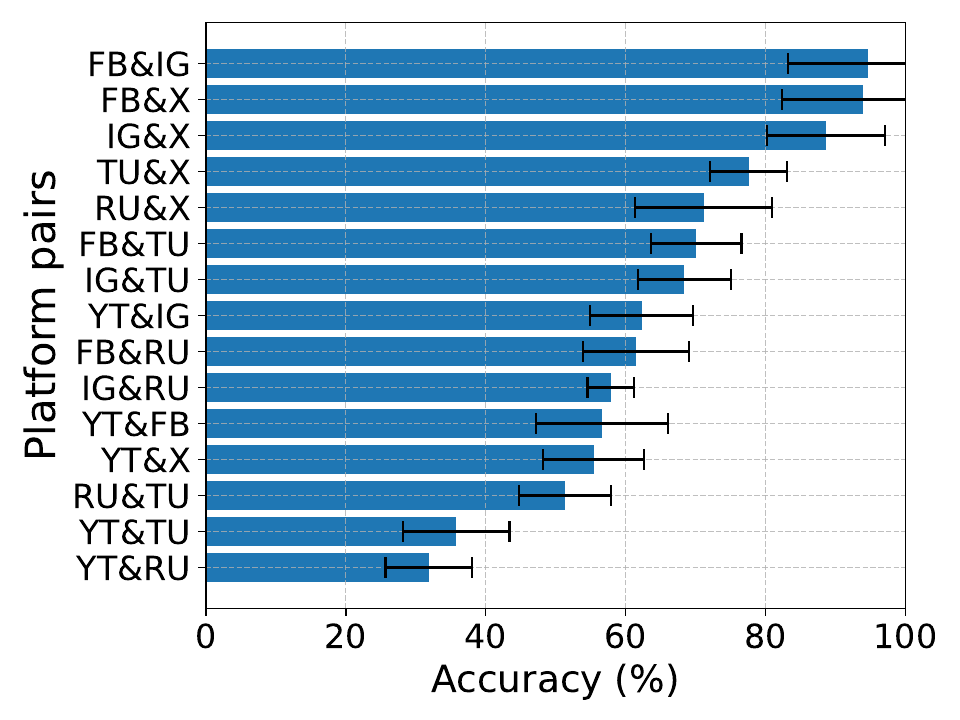}
    \label{fig:Figure13_a}
  } 
    \subfloat[Testing platforms]{% 
    \includegraphics[width=0.19\textwidth]{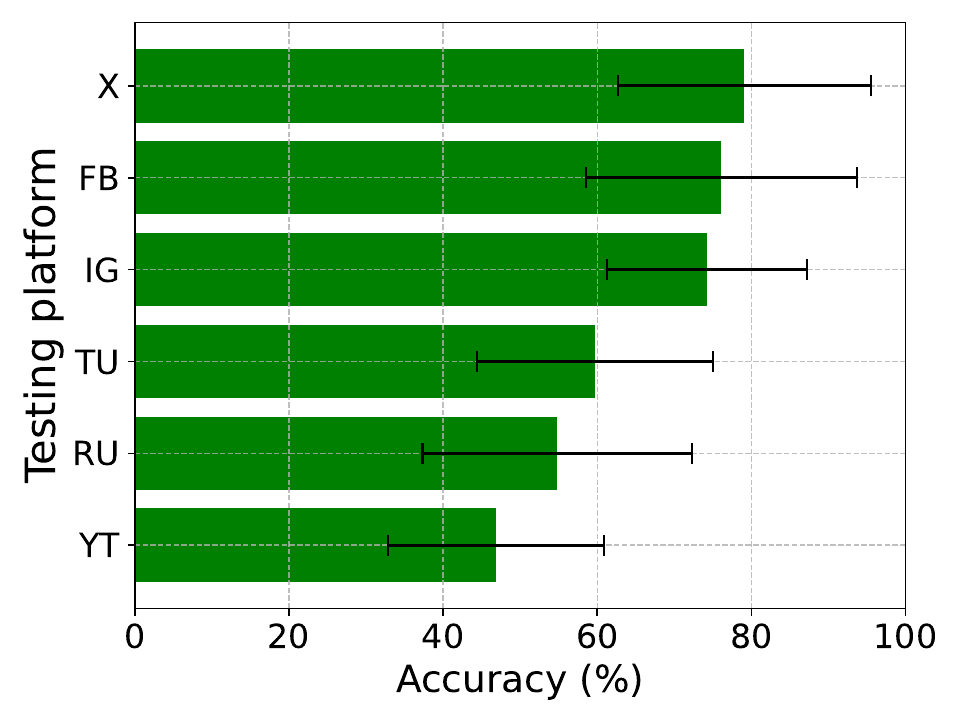}
    \label{fig:Figure13_b}
  } 
    \subfloat[Training platforms]{% 
    \includegraphics[width=0.19\textwidth]{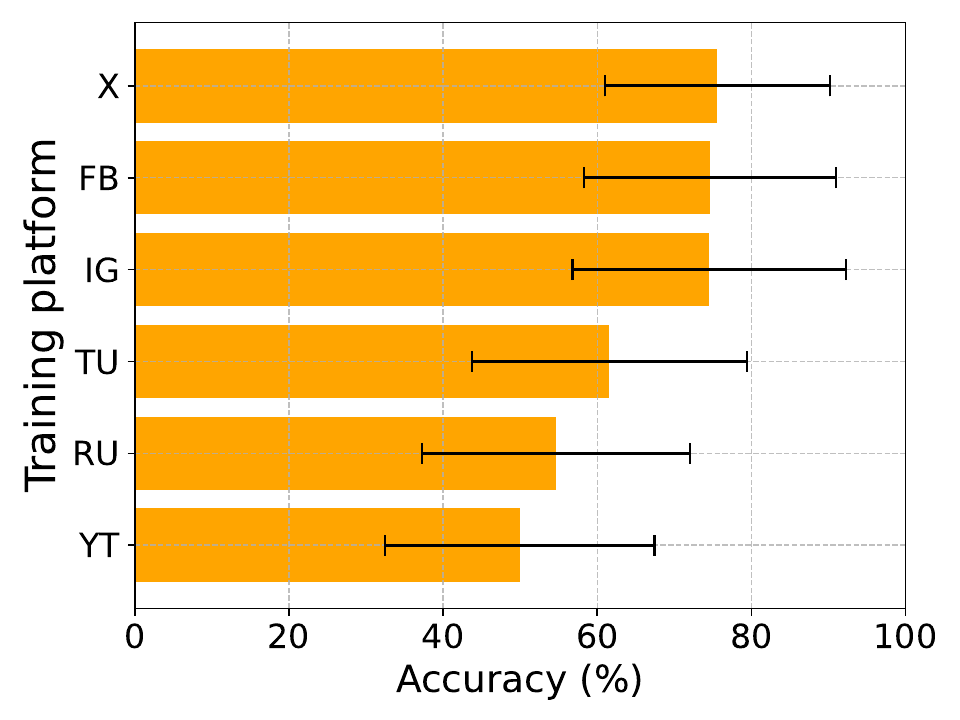}
    \label{fig:Figure13_c}
  } 
  \subfloat[Raw input]{% 
    \includegraphics[width=0.19\textwidth]{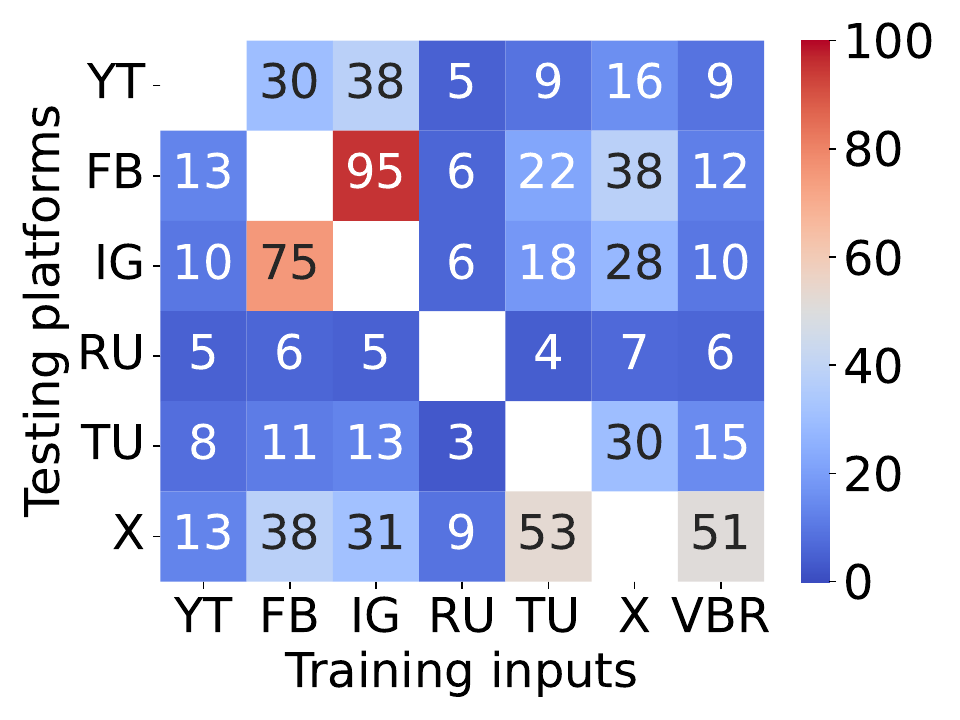}
    \label{fig:Figure11__a_}
  } 
    \subfloat[Embeddings]{% 
    \includegraphics[width=0.19\textwidth]{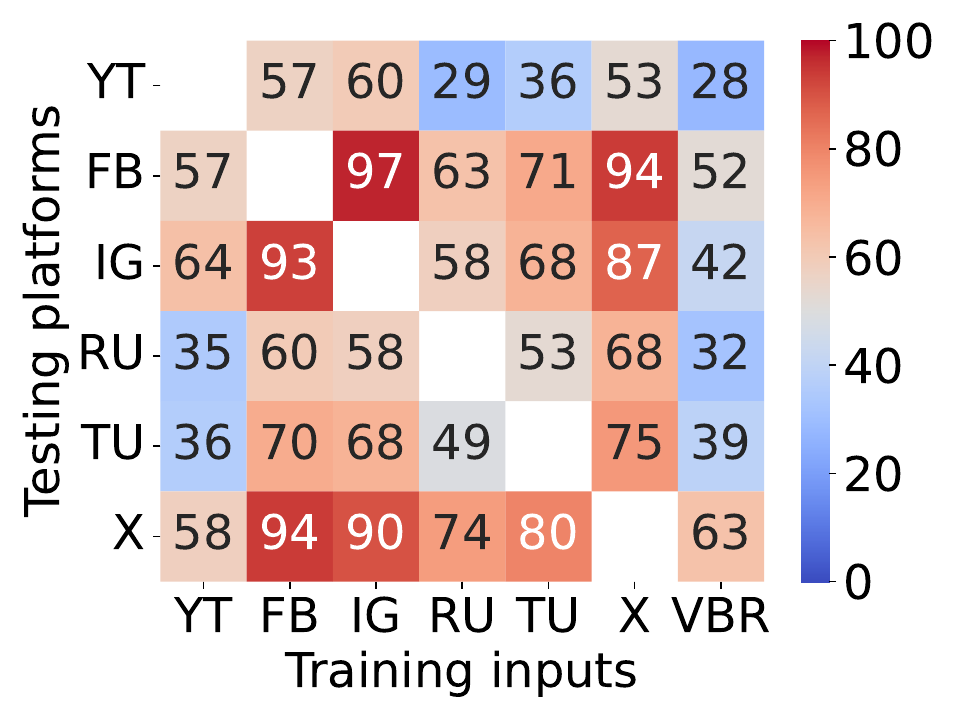}
    \label{fig:Figure11__b}
  }
  \caption{Mean cross-platform accuracies for cross-platform pairs and training/testing platforms for the closed set of 20 videos} \vspace{-2mm}
  \label{fig:Figure13w}
\end{figure*}

\begin{figure}[!h] 
\centering
  % \\
  \subfloat[Binary classification: raw input]{% 
    \includegraphics[width=0.22\textwidth]{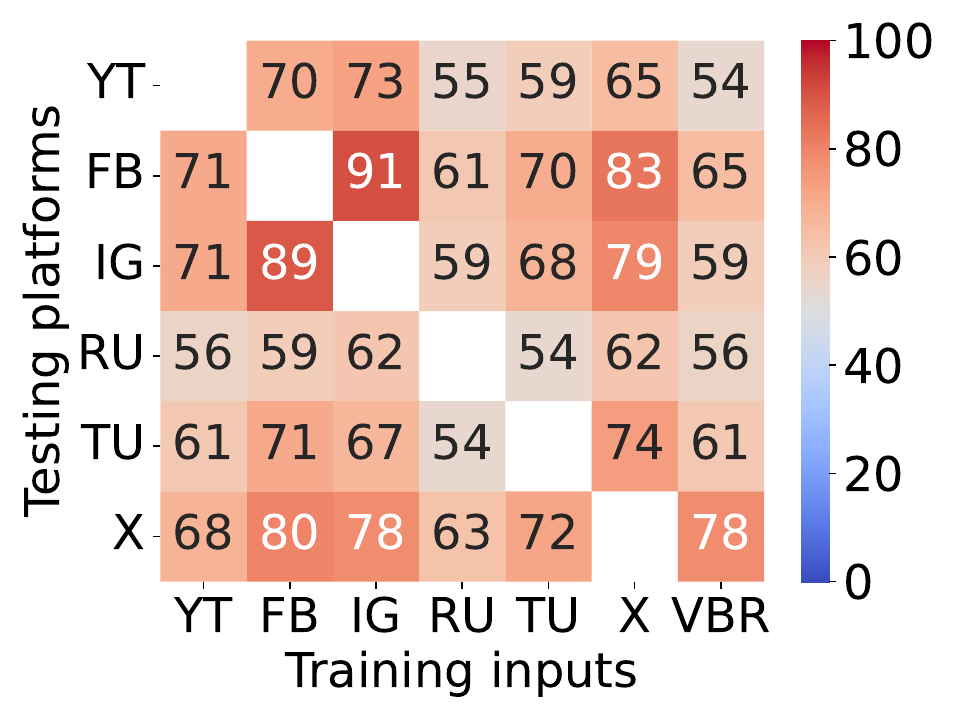}
    \label{fig:Figure11__c}
  } 
    \subfloat[Binary classification: Embeddings]{% 
    \includegraphics[width=0.22\textwidth]{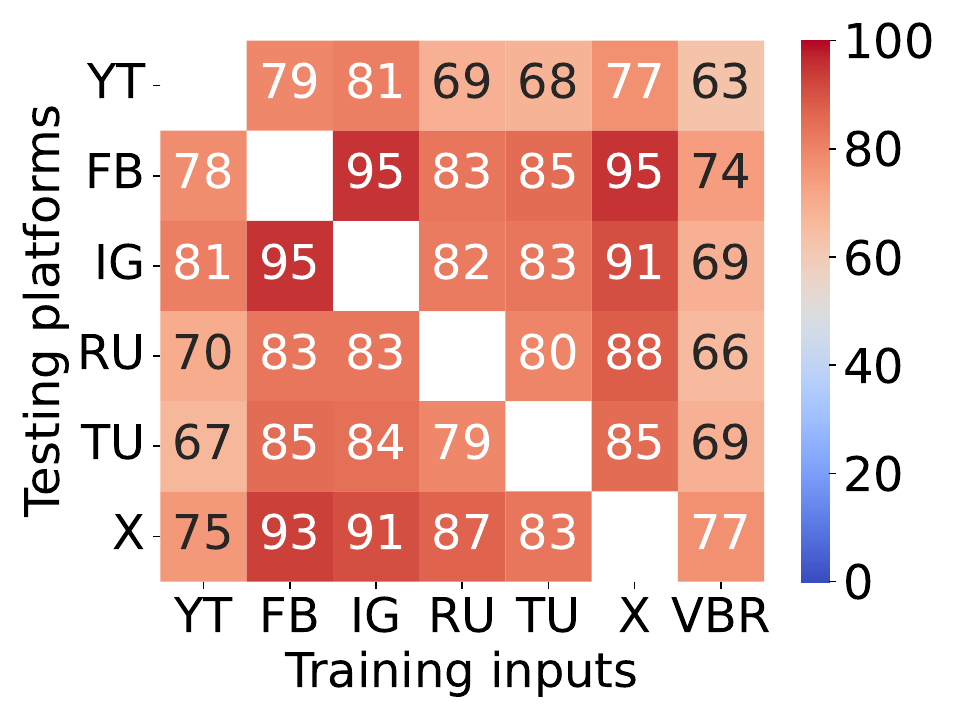}
    \label{fig:Figure11__d}
  } 
  \caption{Accuracy heatmaps for cross-platform binary-class classification on a closed set of 20 videos, before triplet learning (raw input) and after triplet learning (Embeddings)} \vspace{-2mm}
  \label{fig:Figure11_x}
\end{figure}

\subsubsection{Multi-class accuracy heatmaps}
Figure~\ref{fig:Figure13w} (d and e) shows accuracy heatmaps with the percentage accuracies before triplet learning and after triplet learning for the closed set of 20 videos, the training input (platform/VBR) is on the x-axis and the testing platform is on the y-axis. We use the best model from hyperparameter tuning, see section~\ref{chap:hyperpara}. As we can see the cross-platform accuracies before triplet learning are extremely low in general, suggesting cross-platform recognition is generally infeasible without triplet learning, except between Facebook and Instagram which have high initial cross-platform accuracy, which can be explained due to the fact they are from the same parent company Meta and may have very similar video streaming infrastructure and characteristics, which can lead to high similarities between them. As we can see, after triplet learning, we see the accuracies improve significantly across the board, and the cross-platform accuracies between X, Instagram, and Facebook are all between 85-96\% which is very impressive, likely due to the very consistent traffic for these 3 platforms which share high VBR dependence. There are also other platforms that when paired with Facebook, Instagram, or X can provide reasonable cross-platform accuracies such as predicting X from Tumblr which has a mean accuracy of about 80\%, and recognizing X from Rumble with 74\% accuracy. Before triplet learning the mean cross-platform accuracy across all platform pairs is about 20\%, and after triplet learning it is about 65\%, more than 3 times higher accuracy on average than without triplet learning, which is very impressive. 

\subsubsection{Binary-class accuracy heatmaps} \label{chap:binary}

Figure~\ref{fig:Figure11__c} and Figure~\ref{fig:Figure11__d} show the cross-platform classification accuracies before and after embedding generation, respectively, for the binary classifier. The highest accuracies are similar to the multi-class classification scenario without binary, with about 95\% accuracy between Facebook and Instagram as the top accuracies, however, the accuracies of all other cross-platform pairs are also somewhat different and are all approximately above 65\% after triplet learning but which is not very impressive with binary classification since it is not much higher than 50\% which is the point of random guessing. The mean binary classification accuracies before embedding generation are about 68\%, rising to about 82\% after triplet learning which is reasonably high, showing reasonable feasibility. We obtained between 63-77\% in binary classification of pairs of videos from the VBR and each of the 6 platforms pattern which is similar to the literature, which obtained 74\% in classifying Youtube and the VBR pattern, but like it is mentioned in the literature, this is unlikely to be sufficient for open-set classification ~\cite{schuster2017beauty}. 

\subsubsection{Platform analysis}

Figure~\ref{fig:Figure13_a} shows the mean cross-platform accuracies after triplet learning for each bi-directional (regardless of which platform is used for training and which is used for testing) cross-platform pair. As we can see the 3 cross-platform pairs of X, Facebook, and Instagram occupy the top 3 spaces of cross-platform pair accuracies which is not surprising as they are the 3 most similar platforms with very high cross-platform recognition accuracies between them. Platform pairs involving the other 3 platforms are all on the lower end, especially Youtube and Rumble, and the 3 cross-platform pairs of Youtube, Rumble, and Tumblr occupy the bottom 3 spaces of cross-platform pair accuracies. 

Figure~\ref{fig:Figure13_b} and Figure~\ref{fig:Figure13_c} show the best platforms for training and testing, respectively, for cross-platform recognition, where the same platform is used for training and the same is used for testing. As we can see the best training platforms are also the best testing platforms and are X, Facebook, and Instagram which is not surprising. The other 3 platforms are the worst, and surprisingly Youtube is the hardest platform to obtain good cross-platform recognition accuracy when used as a training or a testing platform, possibly due to factors like the traffic being less VBR-dependent. We can also see that X is the best platform to train on to classify other platforms.

\begin{figure}[!h] 
\centering
  \subfloat[Multi-class - raw input]{% 
    \includegraphics[width=0.22\textwidth]{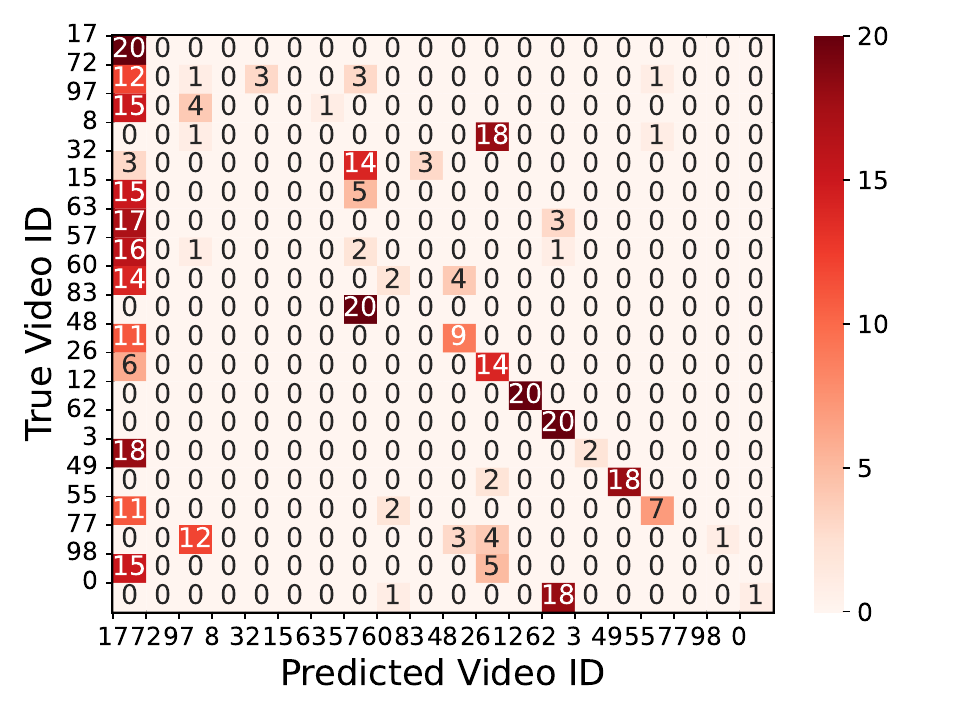}
    \label{fig:Figure15__a}
  } 
  \subfloat[Multi-class - embeddings]{% 
    \includegraphics[width=0.22\textwidth]{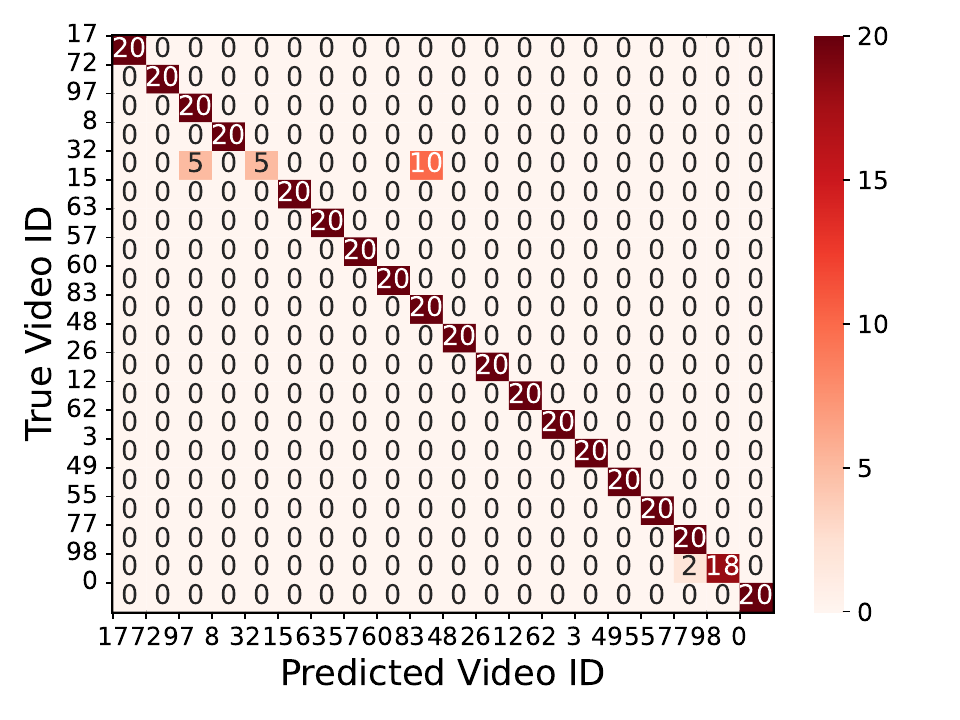}
    \label{fig:Figure15__c}
  } 
\caption{{Multi-class confusion matrices for the closed-set scenario (20 videos) trained on Facebook and tested on X}} \vspace{-2mm}
  \label{fig:Figure15_qq}
\end{figure}
\begin{figure}[!h] 
\centering
  \subfloat[Multi-class - raw input]{% 
    \includegraphics[width=0.22\textwidth]{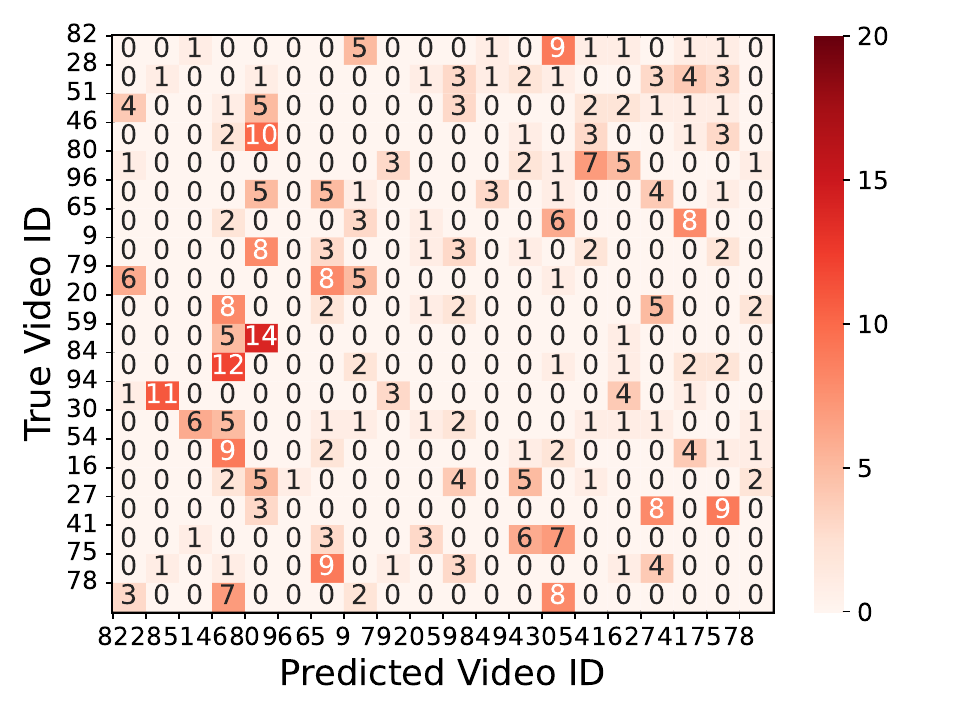}
    \label{fig:Figure15__e}
  } 
  \subfloat[Multi-class - embeddings]{% 
    \includegraphics[width=0.22\textwidth]{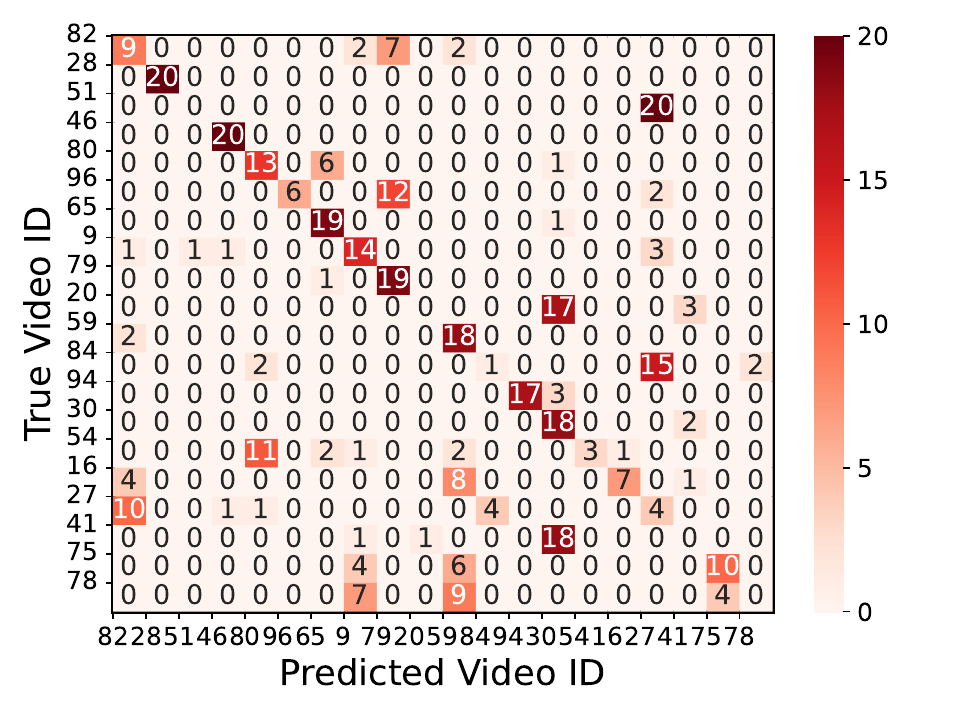}
    \label{fig:Figure15__g}
  } 
  \caption{Multi-class confusion matrices for the closed-set scenario (20 videos) trained on Rumble and tested on Tumblr} \vspace{-2mm}
  \label{fig:Figure15___a}
\end{figure}

\begin{figure*}[!h] 
\centering
  \subfloat[Platform pairs]{% 
    \includegraphics[width=0.19\textwidth]{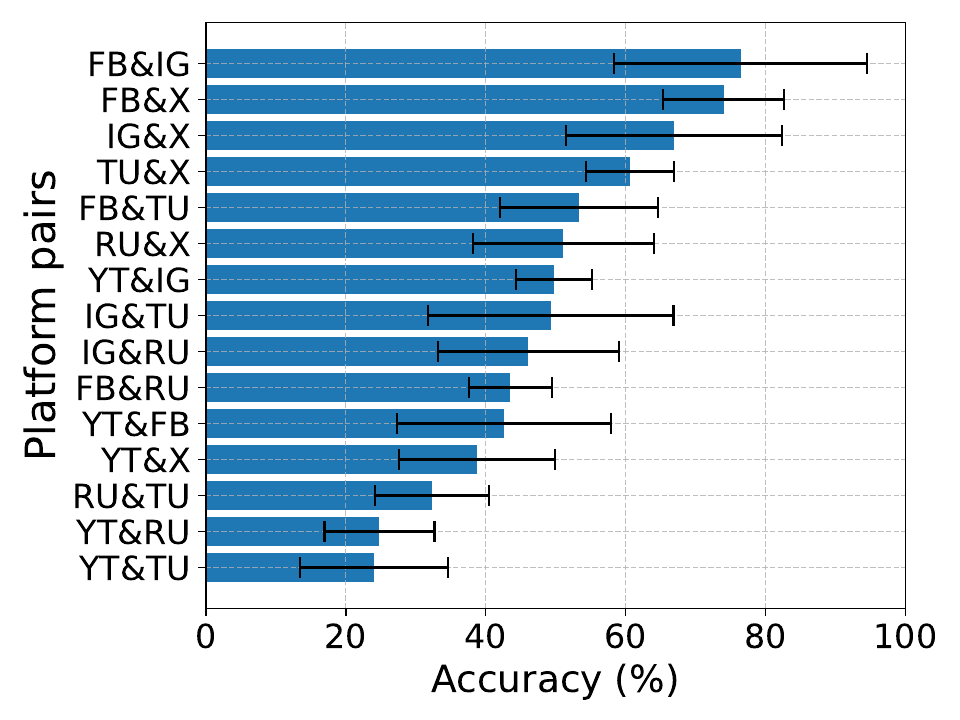}
    \label{fig:Figure32_a}
  } 
    \subfloat[Testing platforms]{% 
    \includegraphics[width=0.19\textwidth]{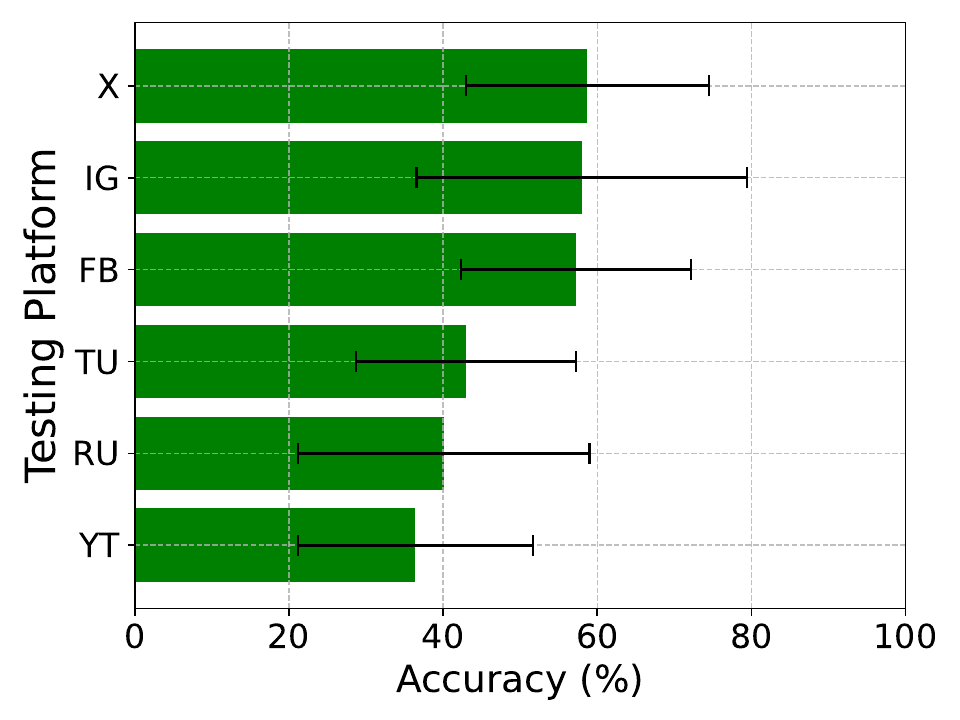}
    \label{fig:Figure32_b}
  } 
    \subfloat[Training platforms]{% 
    \includegraphics[width=0.19\textwidth]{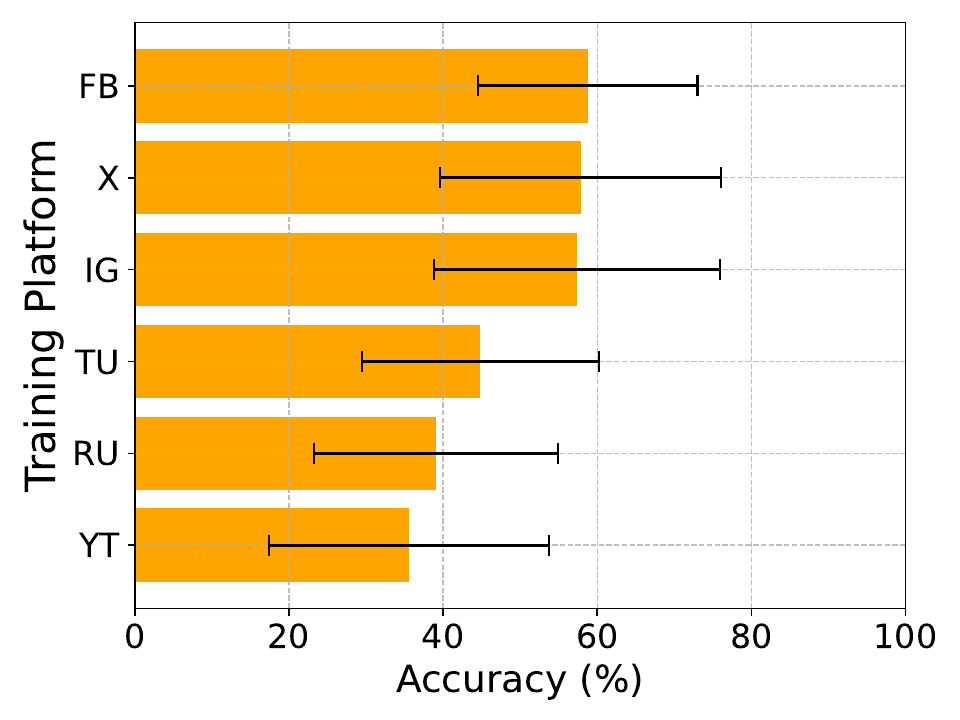}
    \label{fig:Figure32_c}
  } 
  \subfloat[Softmax threshold tuning]{% 
    \includegraphics[width=0.19\textwidth]{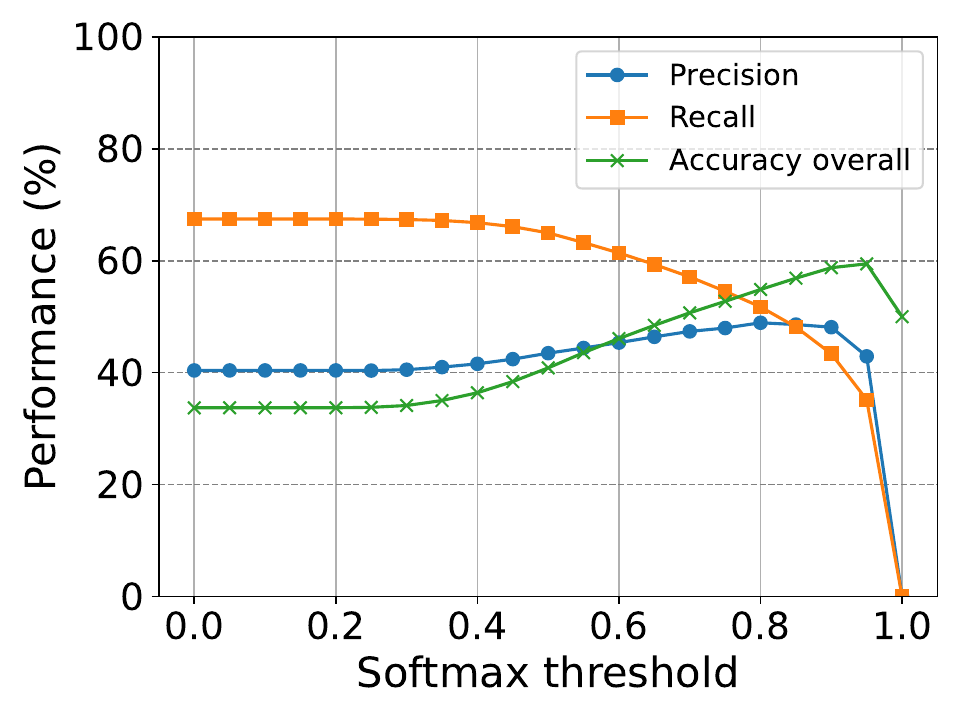}
    \label{fig:Figure27_a}
  } 
    \subfloat[Comparison with closed set]{% 
    \includegraphics[width=0.19\textwidth]{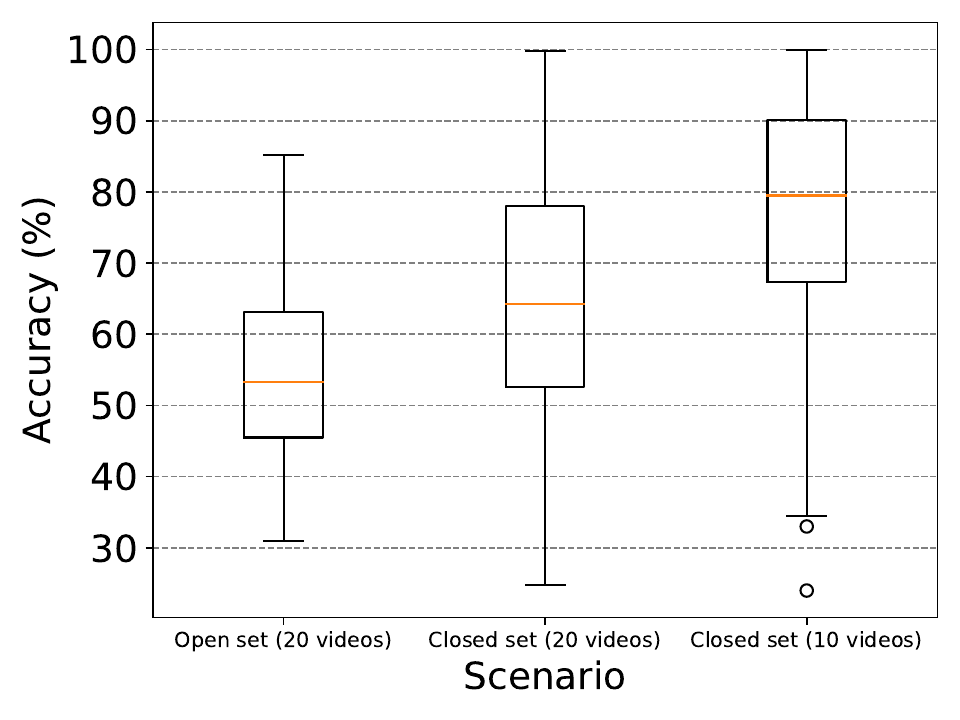}
    \label{fig:Figure27_b}\vspace{-2mm}
  } 
  \caption{Open set cross-platform mean precisions for known videos} \vspace{-2mm}
  \label{fig:Figure32}
\end{figure*}

\subsubsection{Confusion matrices}
Figures~\ref{fig:Figure15_qq} and ~\ref{fig:Figure15___a} show multi-class confusion matrices for single runs of 2 sets of training and testing platforms to offer insight into the impact of different videos/classes on classification performance. From Figure~\ref{fig:Figure15__a} and Figure~\ref{fig:Figure15__c} for multi-class classification, training on Facebook and testing on X, the accuracy before triplet learning is quite poor, with only 3 videos classified perfectly and few videos classified well, and almost all videos are misclassified for video 17. Similarly, from Figure~\ref{fig:Figure15__e} and Figure~\ref{fig:Figure15__g} for multi-class classification, training on Rumble, and testing on Tumblr, we can see the accuracy before triplet learning is very poor. Then after triplet learning, there is a huge difference in both cases, fixing misclassifications and boosting classification performance in general.

\subsection{\framework\ classification performance: Open set}
We tested an open set scenario with 10 seen and 10 unseen videos, using 10 videos to train the CNN with softmax output layer and 20 videos (with 10 additional video classes that were not trained on) in the testing set. We apply softmax thresholding to the probability of each class outputted by the softmax layer of the CNN,  to classify a video as "unknown" (not in the 10 seen videos training set") if the highest probability of any class in the softmax output is below a certain threshold, otherwise, we use the highest probability to get the classified class as normal. We mainly focus on the mean of the precision for each known class. We look at overall accuracy, as well as the mean precision and mean recall of each known class seen at training time, with an approach similar to ~\citet{schuster2017beauty}. However, we focus on mean precision for trained videos to minimize false positives which can be more important in practice, for example, to avoid censorship issues. 

\subsubsection{Softmax threshold tuning}
Here in Figure~\ref{fig:Figure27_a}, we can see the known video precision and recall, as well as the overall accuracy in the open-set scenario for all cross-platform pairs overall, before and after triplet learning. Before triplet learning, the metrics were all very low regardless of the threshold. However, after triplet learning, we see that higher values can be obtained. The highest precision of about 49\% overall can be attained with a threshold of 0.80, the highest recall of about 67\% can be attained with thresholds between 0 and 0.20, and the highest accuracy of about 60\% can be attained with a threshold of about 0.95. Of course for specific platform pairs, the performance of these metrics according to the threshold can vary but we use the best threshold overall for each metric for classifying the unknown class. We also see that the lower the threshold the higher the recall, but the precision and overall accuracy peak close to the threshold of 1 before they drop significantly at 1. 

\subsubsection{Comparing open-set with closed-set scenarios}
Here Figure~\ref{fig:Figure27_b} shows a comparison of the distribution of accuracies in each of 3 scenarios, open set of 20 videos, closed set of 10 videos, and closed set of 20 videos. As we can see, The closed set of 10 videos consistently has much higher performance in general, followed by the closed set of 20 videos and finally the open set. This meets expectations since closed-set scenarios are easier to obtain high accuracies for, and the open-set scenario is the most challenging. We see that the mean accuracy for open-set scenario of 10 seen and 10 un-monitored videos is about 55\%, for the closed set of 20 videos it is about 65\%, and for the closed-set scenario of 10 videos, it is about 77\%, showing a significantly higher accuracy for the closed-set scenarios, especially the closed set of 10 videos. This highlights the challenges with the open-set scenario, for example, perhaps the un-monitored videos might cause significant misclassifications if they are similar to some of the seen videos.

\subsubsection{Platform analysis}
Figure~\ref{fig:Figure32_a} shows the mean cross-platform precisions after triplet learning for each bi-directional (regardless of which platform is used for training and which is used for testing) cross-platform pair. In addition, Figure~\ref{fig:Figure32_b} and Figure~\ref{fig:Figure32_c} show the best training platforms are the best testing platforms and are X, Facebook, and Instagram. The rankings are similar to the closed-set scenarios for accuracy, giving valuable insight into precision in the open set. As we can see the 3 cross-platform pairs of X, Facebook, and Instagram occupy the top 3 spaces of cross-platform pair precisions and the platform pairs involving the other 3 platforms are all on the lower end, especially Youtube and Tumblr. Not surprisingly, not one platform pair attains at least 80\% precision, highlighting how challenging the open-set scenario, even with 10 unseen videos, really is. Nevertheless, 70\% to 80\% can still be considered reasonable in some circumstances, so the results suggest some feasibility in open-set classification. \vspace{-2mm}

\subsection{Impact of varying number of classes and trials}
We also experimented with varying the number of classes used to train the triplet model and also varying the number of classes used for classification, in the closed set. We tried using 20, 40, 60, and 80 classes to train the triplet model whilst using 20 mutually exclusive classes for the classification task, and we also tested the closed set of 10 videos classification task with a mutually exclusive set of 90 videos/classes for training the triplet model. 

\begin{figure}[!h]
    \centering
    \includegraphics[width=0.6\linewidth]{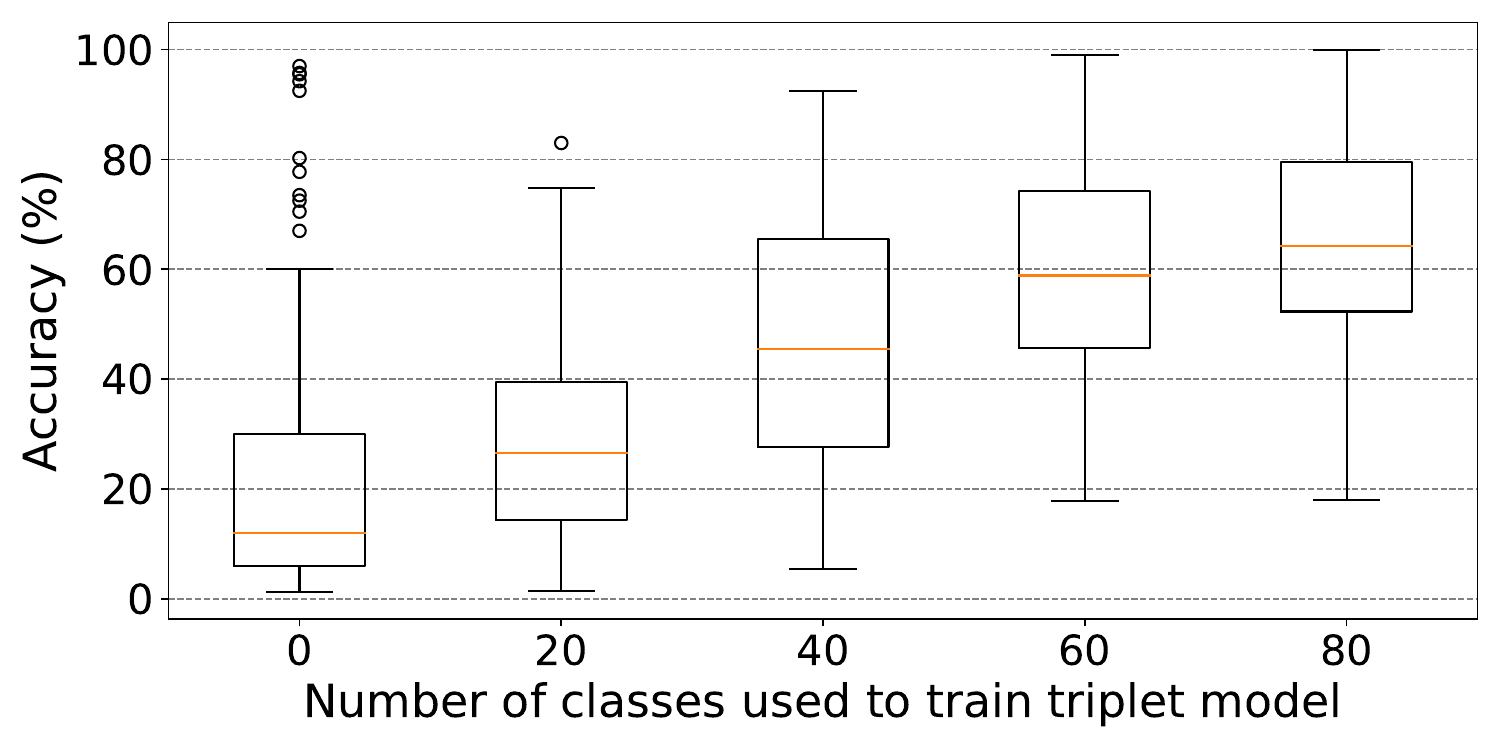}
    \caption{Varying number of training classes}
    \label{fig:Figure38_a}
\end{figure}

\subsubsection{Varying number of training classes}
As we can see from Figure~\ref{fig:Figure38_a}, our results show that the greater the number of classes used to train the triplet learning feature extractor, the higher the final recognition accuracy of the triplet learning model. Without triplet learning the mean accuracy is about 21.4\%, for 20 classes of triplet training it is about 29.3\%, for 40 classes it is about 46.9\%, for 60 classes it is about 59.7\% and for 80 classes it is about 65.2\%, showing improvements with more training classes. Like the ~\citet{web2019} mentioned, the triplet learning model learns to generate better embeddings by learning the differences between classes so having many classes to train the triplet learning model is crucial.

\subsubsection{Accuracy heatmaps for a closed set of 10 videos}

\begin{figure}[!h] 
\centering
  \subfloat[Raw input]{% 
    \includegraphics[width=0.23\textwidth]{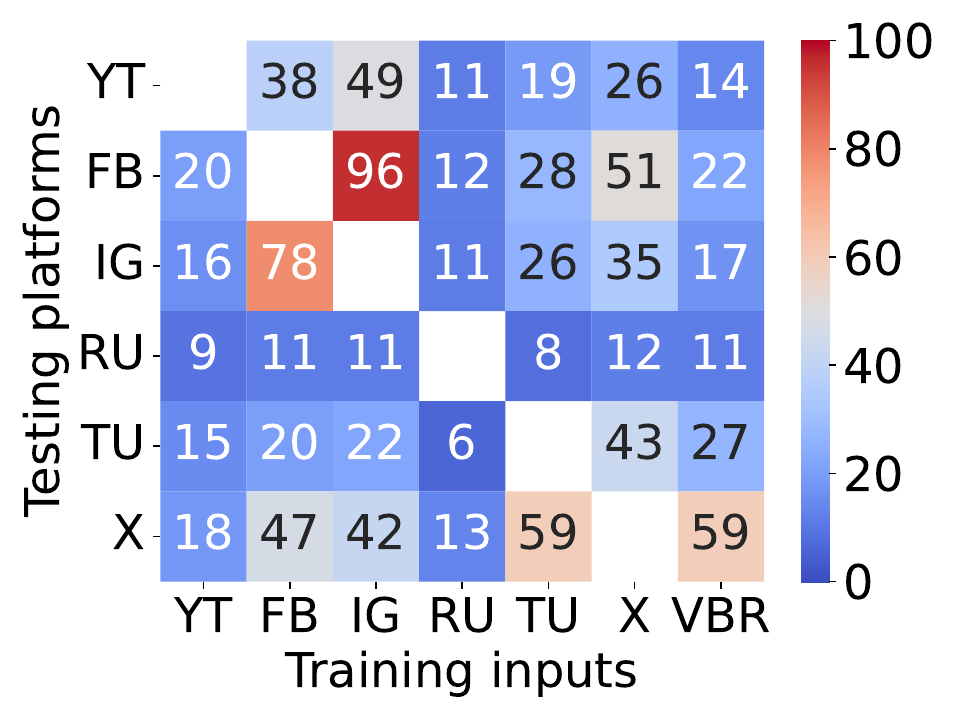}
    \label{fig:Figure11__a}
  } 
    \subfloat[Embeddings]{% 
    \includegraphics[width=0.23\textwidth]{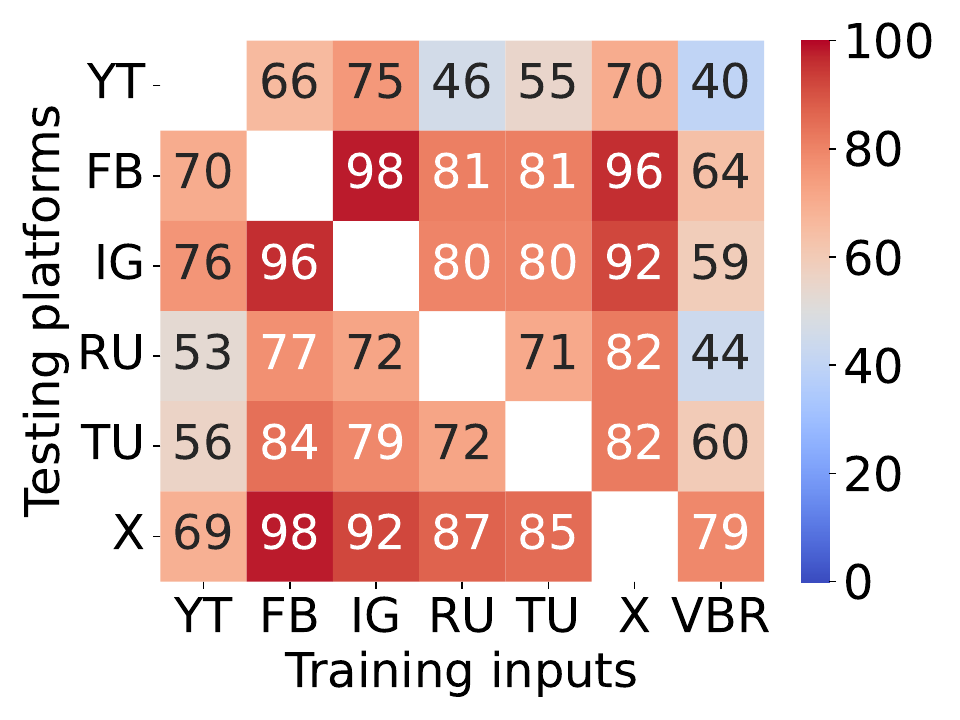}
    \label{fig:Figure11__b_}
  } 
  \caption{Accuracy heatmaps for cross-platform multi-class classification on a closed set of 10 videos, before triplet learning (raw input) and after triplet learning (embeddings)} \vspace{-2mm}
  \label{fig:Figure11_}
\end{figure}

Figure~\ref{fig:Figure11_} shows accuracy heatmaps for the closed set of 10 videos. The mean cross-platform accuracy across all platform pairs is about 28\% before triplet learning which is very low, rising to about 77\%, about 3 times higher with triplet learning, which is very impressive. The final mean accuracy is significantly better than for the closed set of 20 videos classification task which meets our expectations because the closed set of 10 videos scenario has more classes to train the triplet model (90, rather than 80) and fewer classes for the classification task (10, rather than 20), which makes it much less challenging.

After triplet learning, the cross-platform accuracies between X, Instagram, and Facebook are all between 92-98\% which is very high. There are also other platforms that when paired with Facebook, Instagram, or X can provide reasonable cross-platform accuracies such as predicting X from Tumblr which has a mean accuracy of about 80\%, and recognizing X from Rumble with 74\% accuracy. Classification of platforms from the VBR is generally infeasible except for X, with about 80\% after triplet learning for classifying x from the VBR, suggesting a strong correlation between the VBR pattern and X, but the VBR has a weaker correlation with the other platforms. 

\begin{figure}[!h] 
\centering
  \subfloat[Varying number of traces per video/class]{% 
    \includegraphics[width=0.22\textwidth]{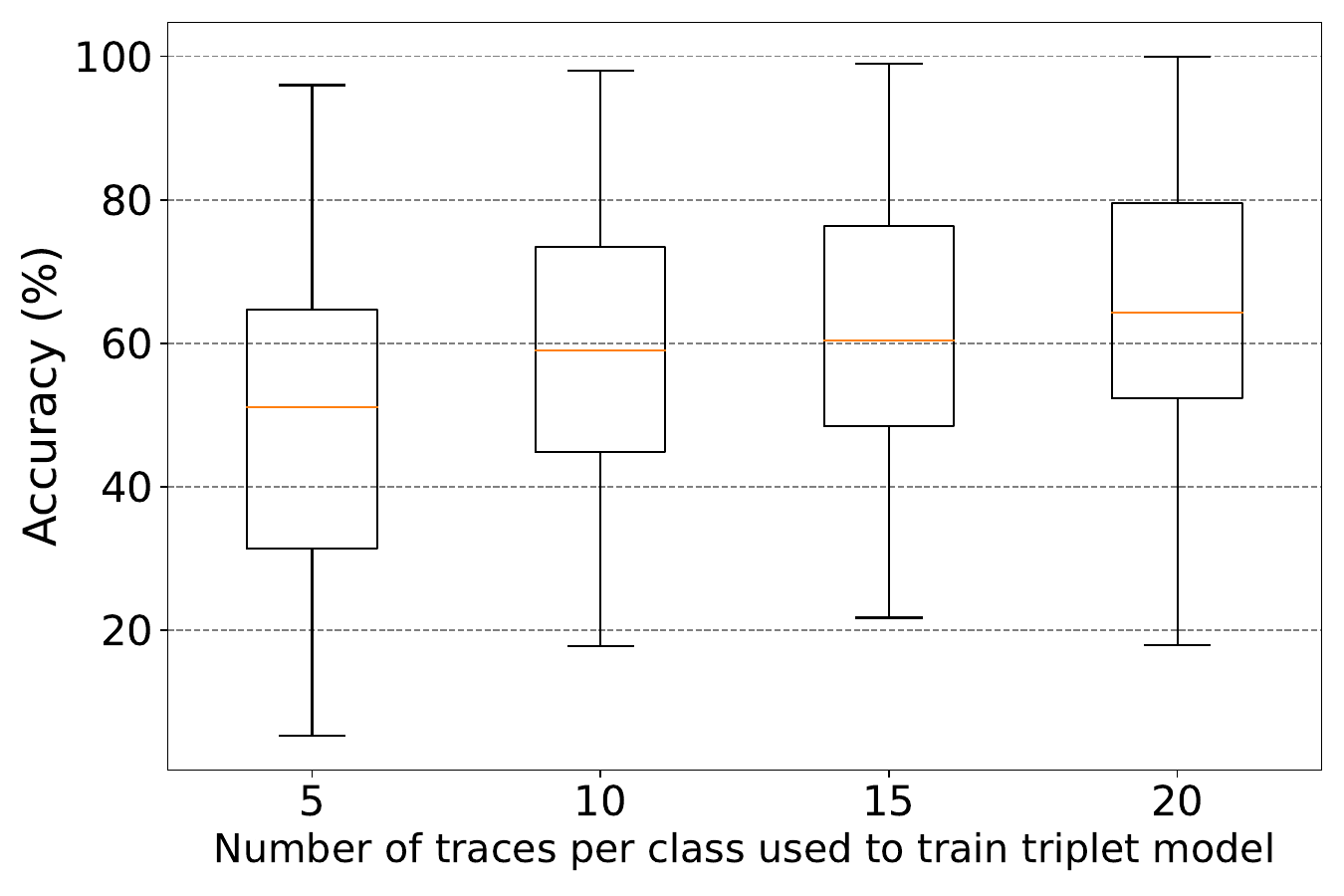}
    \label{fig:Figure3g_a}
  } 
    \subfloat[Data augmentation]{% 
    \includegraphics[width=0.22\textwidth]{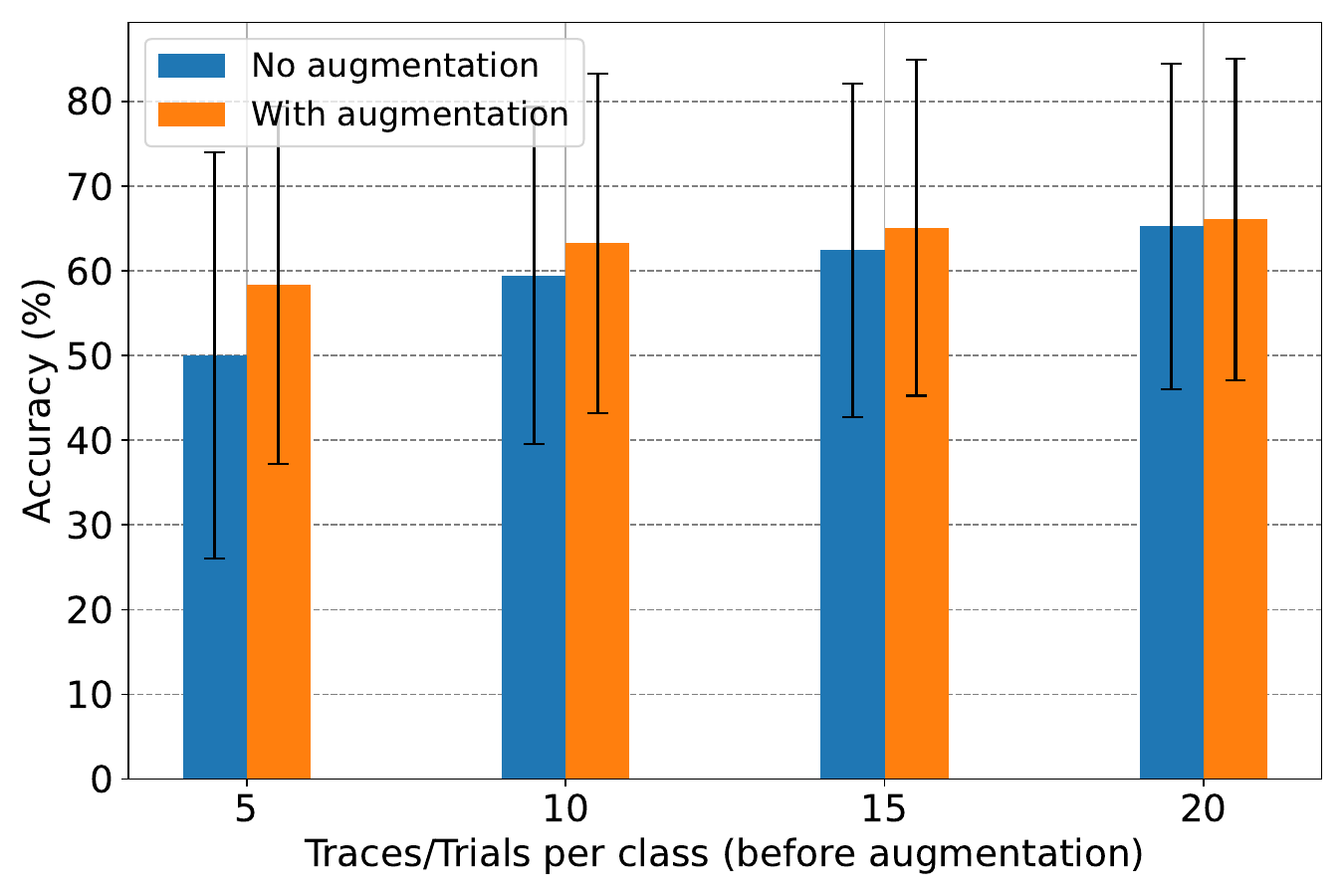}
    \label{fig:Figure36_bh}
  } 
  \caption{Varying number of training trials and data augmentation} \vspace{-2mm}
  \label{fig:Figure36}
\end{figure}

We also conducted tests into varying the number of traces per class used to train the triplet learning model, and we also tried some data augmentation to artificially boost the number of traces per class for different scenarios of the number of traces per class.

\subsubsection{Varying number of traces per video}
We try to vary the number of training trials. We conduct tests with 5, 10, 15, and 20 trials and as we can see from Figure~\ref{fig:Figure3g_a}, we find the performance improvements from increasing the number of classes are not very high and is very small in comparison to the performance improvements from increasing the number of training classes as can be seen in Figure~\ref{fig:Figure38_a}, which can be explained since each video has a fairly stable pattern for the same video, with little intra-class variability ~\cite{web2019}.

\subsubsection{Data augmentation}
In another experiment, we applied data augmentation by adding 5 augmented traces for each class in the training set for the triplet learning, adding Gaussian noise to 5 existing samples. We heuristically tried different augmentation techniques and decided upon sampling from a Gaussian distribution with a mean equal to the values in a traffic sample vector, and a standard deviation equal to 5\% of the value of each element. As Figure~\ref{fig:Figure36_bh} demonstrates, the results show that when the original training set has only 5 samples, the additional 5 augmented samples improve accuracy substantially, to a similar accuracy as for the 10 normal traces per class and 15 normal traces per class. For a higher number of trials per class, the data augmentation does not have a significant improvement, again perhaps due to limited intra-class variability.

\subsection{Robustness to video modification}
\begin{figure}[!h]
    \centering
    \includegraphics[width=0.50\linewidth]{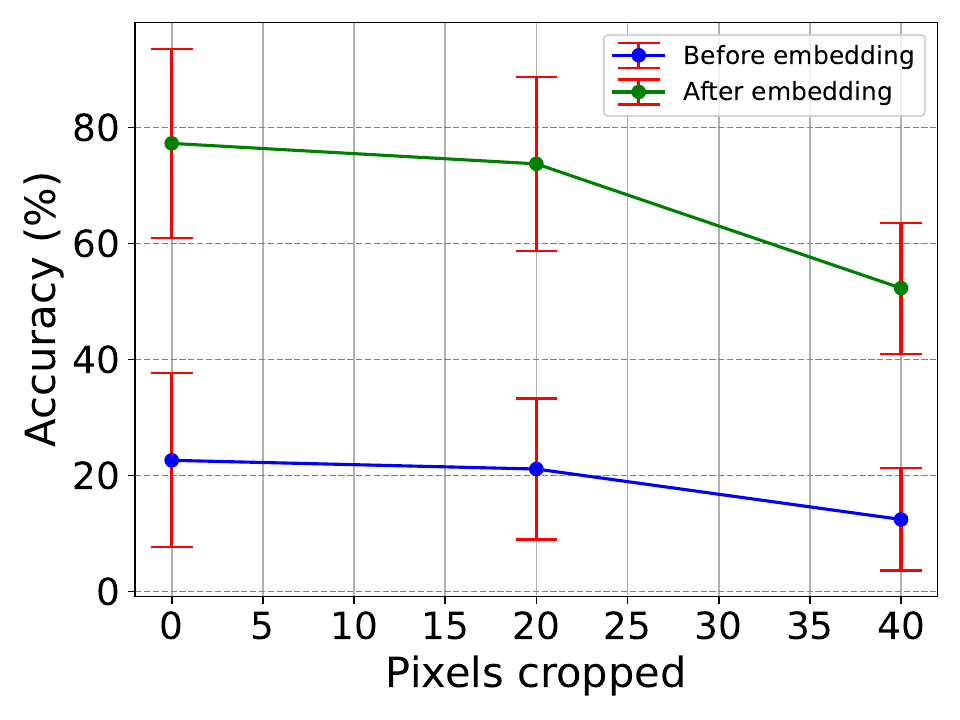}
    \caption{Mean cross-platform recognition accuracy of recognizing videos on X using the encrypted traffic of the same videos on other platforms, with a varying amount of modification/cropping of pixels of the testing videos on X} \vspace{-3mm}
    \label{Figure 43}
\end{figure}
Major platforms like Google/YouTube, Facebook, and X can detect re-uploads of copyrighted videos, illegal videos, and harmful misinformation videos even with slight modification, with tools such as robust hashing algorithms that calculate the hash for a video and compare it to a database of hashes of flagged videos ~\cite{kbzkhashingCould}. However, even these methods are not foolproof and can miss many instances ~\cite{kbzkhashingCould}. Attackers use a larger variety of attacks and the attack we decided to test the robustness of our method with was cropping attacks, which are considered one of the most prominent ~\cite{SALIENT2010}.

Our method also has application as a video copy detection technique which can be applied to identify pirated movies and other video copies violating copyright and being unlawfully uploaded to platforms. We used a similar approach to this work, ~\citet{SALIENT2010}, where we cropped the video 40 pixels from the top and 40 from the bottom, and we also used 20 pixels instead of 40 as an additional test, and we uploaded the 100 videos with these modifications to X and collected the traffic. Some of our videos have pixel dimensions of 864x540 whilst others have 960x540 so cropping 40 pixels from them would result in ratios of 864x500 and 960x500 respectively. 

The results in Figure~\ref{Figure 43} show that with 20-pixel cropping the accuracy is still very high, similar to without any cropping, but then for 40-pixel cropping, the accuracy suddenly drops significantly. This shows that our method is robust to a small or reasonable amount of modification to the video but its effectiveness can be significantly reduced and possibly even broken by larger modifications. However, our method is likely to be more robust than many other techniques such as hashing because small modifications like the insertion of a frame do not really change the VBR of the video, if at all, which keeps the VBR pattern stable and allows our method to maintain high recognition accuracy when other methods may not.

\subsection{Further analysis: Hyperparameter tuning} \label{chap:hyperpara}
\begin{figure}[!h] 
\centering
  \subfloat[Different bin sizes and models]{% 
    \includegraphics[width=0.23\textwidth]{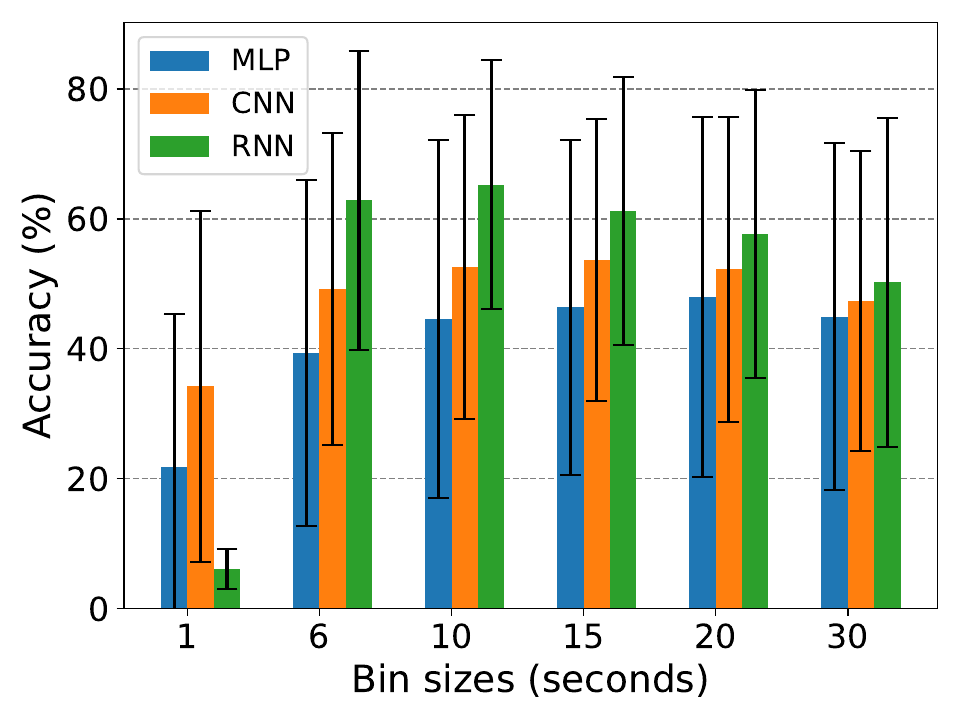}
    \label{fig:Figure10_a}
  } 
    \subfloat[Different traffic durations]{% 
    \includegraphics[width=0.23\textwidth]{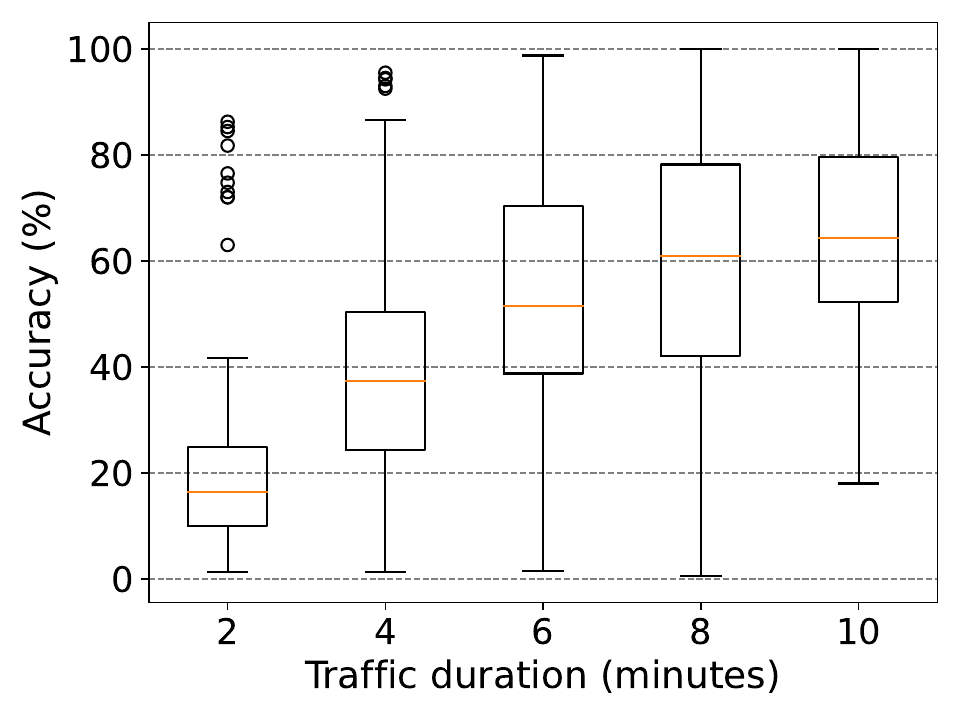}
    \label{fig:Figure10_b}
  } 
  \caption{Hyperparameter tuning} \vspace{-3mm}
  \label{fig:Figure10}
\end{figure}

\subsubsection{Comparing bin sizes and triplet base models}\label{sec:hyperp_basemodels}
Figure~\ref{fig:Figure10_a} shows results for different bin sizes and base model architectures for the triplet model. Here we see that the small bin sizes used in the literature which work well to identify videos within a single platform have by far the worst performance, suggesting the noise and varying segmentation parameters between the platforms may be contributing factors. ~\citet{afandi2022fingerprinting} argue that aggregating the bytes into larger bins (6 seconds) helps deal with the irregularities and inconsistencies that the VBR encoding causes, however, we see that for cross-platform recognition we need even larger bin sizes like between 10-20 seconds for optimal performance. Yet, for 30 seconds the accuracy drops significantly, indicating 30 seconds and higher may be too coarse-grained, reducing the amount of detail and features in the data, and reducing the ability to recognize videos across platforms. As we can see a bin size of about 10 seconds and the RNN method seems optimal. 

RNN may be superior to the other methods since it is good at dealing with time series data, and CNN superior to MLP because it is more robust to variability of the data for small bin sizes. It shows that the 20-second bin size is optimal for MLP with a best mean accuracy of about 48\%, the 15-second bin size is optimal for CNN with a best mean accuracy of about 53.6\%, and the 10-second bin size is optimal for RNN with a best mean accuracy of about 65.2\%. We assumed the best model and bin size for 10 minutes was optimal for all other durations as well in general, so we took the best model and bin size for 10 minutes, and evaluated its performance for different durations as can be seen in Figure~\ref{fig:Figure10_b}. 

\subsubsection{Comparing traffic durations}
Figure~\ref{fig:Figure10_b} shows a comparison of cross-platform classification accuracy distributions for different traffic capture durations for the best model of 10-second bin size RNN. It shows that longer durations provide better accuracies in general. We can see some outliers for 2 and 4-minute traffic durations with high accuracies which are mainly Facebook Instagram and x. For 2 minutes, recognition is only feasible between Facebook and Instagram but they are already expected to be similar anyway, as they are both from the same parent company Meta and therefore likely share the same video streaming technology and infrastructure. With 4 minutes, recognizing using X is also reasonably feasible. From 6 minutes and higher most other platforms are also feasible. However, recognizing videos with traffic durations below 6 minutes is very challenging in general. This is the reason why we used 10 minute traffic traces, because the task is more challenging for the shorter traces used in previous literature. 

The Figure~\ref{fig:Figure10_b} shows a fairly steady but decreasing rise in accuracy with longer duration, which indicates that even longer durations may give even higher performance but 10 minutes approaches the limit of what is practically and realistically feasible. The mean cross-platform accuracy for 2 minutes is about 20.9\%, for 4 minutes it is about 39.7\%, for 6 minutes it is about 51.1\%, for 8 minutes it is about 58.4\% and for 10 minutes it is about 65\%. The best accuracy is obtained with 10 10-minute duration, which makes sense because more data allows for more common patterns among the same video on different platforms to emerge, allowing for more similarities and differences between videos of the same and different classes. However, 10 minutes is also very long showing a limitation for this approach. Moreover, we assume since there is a strong positive correlation between duration and recognition accuracy for the best model and bin size, we assume this applies to all models and bin sizes in general. 

\subsection{Further analysis: classifiers} \label{chap:classifiers}
\begin{figure}[!h] 
\centering
  \subfloat[Accuracy distributions]{% 
    \includegraphics[width=0.23\textwidth]{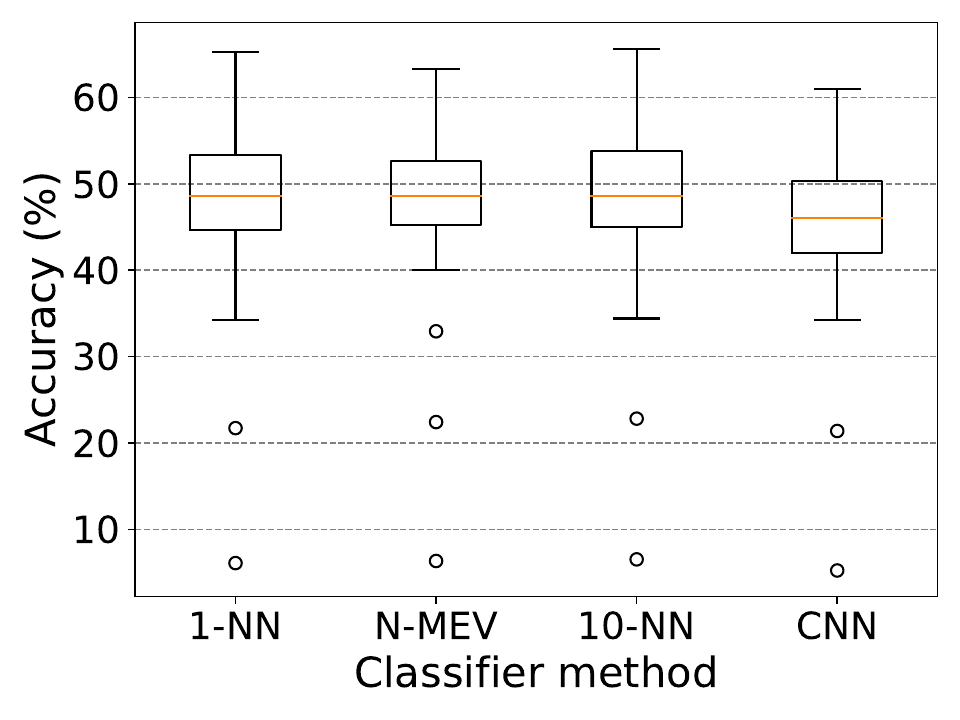}
    \label{fig:Figure36_a}
  } 
    \subfloat[N-shot learning]{% 
    \includegraphics[width=0.23\textwidth]{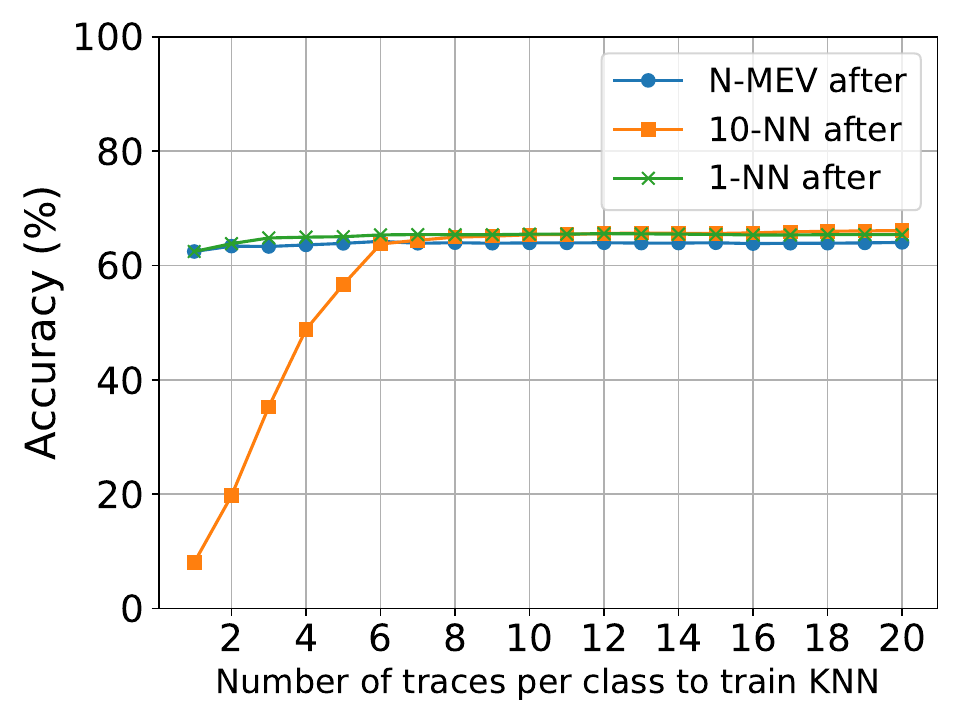}
    \label{fig:Figure36_b}
  } 
  \caption{Comparison of classifier accuracy distributions for the hyperparameter tuning (a) and comparison of KNNs for N-shot learning (b)} \vspace{-3mm}
  \label{fig:Figure36_}
\end{figure}

\subsubsection{Comparing accuracy of classifiers}
We used KNN (all using Euclidean distance) with 1 and 10 nearest neighbours, as well as the N-MEV approach from the paper ~\citet{web2019}, and we also tried a neural network classifier. As can be seen from Figure~\ref{fig:Figure36_a}, we found that the KNNs consistently provided the best performance and the "N-MEV" approach seemed to have a distribution concentrated on higher accuracies compared to others. The mean accuracy of each KNN is around 46\% and for the CNN it is about 44.15\%, so the KNNs are superior. 

\subsubsection{Comparing KNNs in N-shot learning}
From Figure~\ref{fig:Figure36_b}, we see that the 10-NN is quite unstable as with less than about 10 trials it attains much lower accuracy than the other techniques, but actually achieves slightly higher performance than the other KNNs for a large number of traces per class. ~\citet{web2019} show that the "N-MEV" approach is a bit unstable and has low performance for a few trials per class such as less than 5, and then plateaus at about 15 trials for N-shot learning, however, when we compare the N-MEV approach, we see that the NMEV approach performs well even with 1 sample and any amount of samples. Based on our analysis of these results, since 1-NN is stable and has high performance, we select it as our main classifier.

%% file: Sections/Chapter6.tex
%-------------------------------------------------------------------------------
\section{Discussion and limitations}\label{chap:Discussion}
%-------------------------------------------------------------------------------

Our study has demonstrated the practical feasibility of \framework\ as a supplementary tool for content moderation, especially in the presence of weak platform-level content moderation or streamed via malicious platforms. 
Moreover, \framework\ has other potential applications that require content identification through passive observations including in digital forensics, content filtering in enterprise networks and parental control in home networks. Furthermore, our approach of using triplet learning to separate out dataset-specific from class-specific features can be applied in diverse domains and applications such as speech recognition across accents and facial recognition across different ages. However, despite these promising applications, our framework \framework\ does have the following limitations that can be improved in future research.

Our cross-platform video recognition approach generally requires long video traces (5-10 minutes) which may not be realistic and practical in real-time detection. Future work could include optimizing the model for shorter and more practical trace durations whilst also maintaining high accuracy. In addition, increasing the amount of training data could also boost performance for shorter durations. 

Our method's effectiveness can also be affected by many factors such as modifications to uploaded videos. While we demonstrate the robustness to some basic modifications, \framework's performance with advanced modifications needs to be further studied. Moreover, user interaction (for example, pausing video), caching, network changes and the use of public videos, can also introduce unexpected variations in the traffic stream. Some of these factors have been explored in the single platform case \citep{kattadige2021seta++,Dubin_2017,amjad2022encvidc}, but future studies should test these variables in the cross-platform case to explore methods to improve robustness. For example, similar works in the single platform case have demonstrated the ability to recognize videos with very high accuracy even in different scenarios of starting point and trace duration \cite{kattadige2021seta++,hassan2022youtube}, which means it may be possible for the cross-platform case as well, and should be explored.  

The open set performance and the performance validation with data collected over 6 months demonstrated the robustness of our method against dynamic changes of platform and network configurations. However, streaming protocols and parameters can change over time ~\cite{schuster2017beauty} and could lead to model invalidation and requirements for model re-training and updates. Future studies could focus on addressing these challenges, perhaps with adaptive learning algorithms to better adapt the model to platform changes more smoothly. 

%-------------------------------------------------------------------------------
\section{Conclusion}\label{chap:conclusion}
%-------------------------------------------------------------------------------

In conclusion, \framework\ has demonstrated the ability to significantly improve the accuracy of recognizing video titles across platforms, especially across Facebook, Instagram, and X which have very VBR-dependent traffic. Whilst our approach shows promise, with a boost in cross-platform recognition accuracies of about 2 or 3 times on average in closed set scenarios and precisions of up to about 80\% in the open set, it is less effective but still in some cases feasible, for platforms like Rumble, YouTube, and Tumblr due to various factors like having less VBR-dependent traffic. The study highlights the infeasibility in general for cross-platform recognition without triplet learning, and the methods to overcome these challenges, laying the foundation for future work. This could include exploring a wider range of platforms, experimenting with other metric learning techniques, and expanding to other applications such as anomaly detection. Further improvements in these areas could enhance the effectiveness of the framework for combating the cross-platform spread of misinformation videos.